\documentclass[
	fontsize=10pt,          
	numbers=noenddot,    	
    parskip=half,        	
    listof=totoc,        	
    bibliography=totoc,  	
	headsepline=true,       
	footsepline=false, 		
    DIV=12                	
]{scrartcl}

\usepackage[utf8]{inputenc}
\usepackage[T1]{fontenc}
\usepackage[english]{babel}

\usepackage{graphicx}

\usepackage{amsmath, amsfonts, amssymb, bbm, bm, mathtools}

\usepackage[numbers]{natbib}

\usepackage[hidelinks, breaklinks]{hyperref} 
\usepackage{url}
\hypersetup{%
  colorlinks=true,
  linkcolor=blue!50!black,
  urlcolor=blue!70!black,
  citecolor=blue!70!black
}

\usepackage[doublespacing]{setspace} 

\usepackage{pifont}   

\usepackage{wrapfig}

\usepackage{nicefrac}

\usepackage{pdfpages}

\usepackage{rotating}

\usepackage[ruled,vlined]{algorithm2e}  

\usepackage{scrlayer-scrpage}
\clearscrheadings                   
\automark{section}                  
\ihead{Comparing Methods for CATE Estimation in Competing Risks}				
\ohead{\pagemark}					
\ifoot{}							
\ofoot{}							

\usepackage{booktabs} 	

\usepackage{csquotes}

\usepackage{booktabs}
\usepackage{caption}
\usepackage[export]{adjustbox} 
\usepackage[left=1truein,right=1truein,top=1truein,bottom=1truein]{geometry}
\usepackage{abstract}

\usepackage[graphicx]{realboxes}
\usepackage{pdflscape}

\usepackage{multirow}

\usepackage{makecell}

\usepackage{tikz}
\usetikzlibrary{positioning}
\usepackage[capitalise]{cleveref} 
\usepackage{dsfont} 
\usepackage{listings}
\usepackage{xcolor}

\definecolor{codebg}{RGB}{235,242,250} 
\definecolor{codeblue}{RGB}{0,0,220} 
\definecolor{codedarkgreen}{RGB}{0,120,0} 
\definecolor{codegray}{RGB}{100,100,100}

\lstdefinelanguage{R}{
  keywords={
    if, else, repeat, while, function, for, in, next, break,
    TRUE, FALSE, NULL, NA, NA_integer_, NA_real_, NA_complex_, NA_character_,
    library, require, return
  },
  otherkeywords={<-, <<-, ~, $, @, :, ::, :::},
  sensitive=true,
  comment=[l]{\#},
  morestring=[b]",
  morestring=[b]'
}
\lstdefinestyle{Rstyle}{
  language=R,
  backgroundcolor=\color{codebg},
  basicstyle=\ttfamily\normalsize,
  keywordstyle=\color{codeblue}\bfseries,
  stringstyle=\color{codedarkgreen},
  commentstyle=\color{codegray}\itshape,
  numberstyle=\tiny\color{codegray},
  showstringspaces=false,
  breaklines=false,
  frame=single,
  rulecolor=\color{codebg},
  tabsize=2,
  columns=fullflexible,
  basewidth=0.55em,
  aboveskip=1.5em,
  belowskip=1.5em,
  alsoletter={.},
  morekeywords=[2]{c,specify_data,generate_data,ifelse,as.factor,cr_cate,data,set.seed}
}

\newcommand{\expval}[1]{{\mathbb E}\left[ #1 \right]} 
\newcommand{\cindep}{\perp \!\!\! \perp} 

\usepackage{mathabx}

\begin{document}

\thispagestyle{empty}
\newgeometry{top=2.5in,bottom=1in,left=1in,right=1in}

\vspace{15em}

\begin{center}
\Large{\textbf{A Guide to Estimating Conditional Average Treatment Effects in Competing Risks Settings}}    
\end{center}

\vspace{5em}

\begin{center} 
    Daniel Klippert$^{1,2}$, Sarah Friedrich$^{3}$, Markus Pauly$^{1,2}$
    
    \vspace{2em}
    
    \footnotesize{
        $^{1}$Department of Statistics, TU Dortmund 
        University\\
        $^{2}$Research Center Trustworthy Data Science and Security, University Alliance Ruhr (UA Ruhr)}\\
        $^{3}$ Institute for Mathematics,  University of Augsburg 
\end{center}

\vspace{4em}
\begin{center}
    \textbf{Autor Note}

Daniel Klippert \includegraphics[height=10pt]{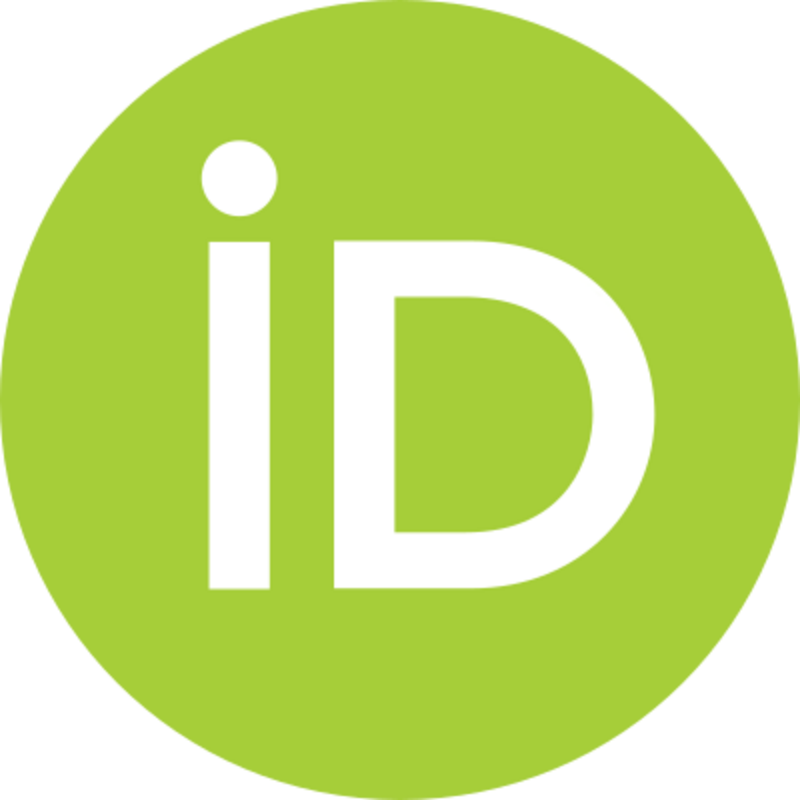} \url{https://orcid.org/0009-0003-3823-1680} \\
Sarah Friedrich \includegraphics[height=10pt]{orcid_logo.png} \url{https://orcid.org/0000-0003-0291-4378}\\
Markus Pauly \includegraphics[height=10pt]{orcid_logo.png} \url{https://orcid.org/0000-0002-0976-7190} 
\vspace{2em}

This study only uses artificially generated (simulation) data, and an openly published data set. The code including all simulations is attached as supplementary material.
We have no conflicts of interest to disclose.
The paper is based on the Master's thesis of the first author, supervised by the other co-authors, and has been substantially extended.
\\

Correspondence concerning this article should be addressed to Daniel Klippert, Department of Statistics, TU Dortmund University, 44227 Dortmund Germany. Email: daniel.klippert@tu-dortmund.de
\end{center}

\newpage

\restoregeometry 

\begin{abstract}
Conditional average treatment effects (CATEs) are central to treatment decision-making in personalized medicine. 
In competing risks settings, estimating CATEs from survival data allows for patient-specific assessments of treatment effectiveness 
for a specific event of interest while properly accounting for alternative event types. This distinction is essential in the presence of comorbidities, where competing causes of death may otherwise confound the therapeutic benefit. Focusing on right-censored survival times with binary treatment, we examine CATEs defined as covariate-conditional differences in the absolute risk for the event of interest at a fixed time. To this end, we study meta-learners which adapt machine learning algorithms for CATE estimation in competing risks scenarios. We systematically compare six meta-learners, combining Cox regression or random survival forests for risk modeling with elastic net regression or random forests for direct CATE modeling. To provide practical guidance on model selection, we evaluate their performance in multiple simulation settings, that differ in hazard complexity, treatment heterogeneity, treatment assignment, event type distribution and censoring. To facilitate applied use, we provide the R package, \texttt{crsurvlearners}, which implements all considered approaches.

\hspace{10mm} \textit{Keywords}: Causal Inference, Survival Analysis, Meta-Learner, Survival Forest, Cox-Regression 
\end{abstract}
\newpage

\section{Introduction} 
Heterogeneous treatment effects (HTEs) capture how the effectiveness of treatment varies across patients with different characteristics \citep{kent:2018}. In practice, HTEs are typically expressed through conditional average treatment effects (CATEs), which relate treatment effects to patient covariates and thereby enable decision-making at the individual level.
CATE estimation is challenging even in standard settings, as it requires isolating causal effects from statistical associations. 
While randomized controlled trials (RCTs) can identify causal effects by design, they are often infeasible due to ethical or practical reasons \citep{gianicolo:2020}. Moreover, they are typically designed to estimate average treatment effects at the population level \citep{longford:1999} and might not include sufficiently diverse populations to support individualized treatment recommendations due to restricted sample sizes \citep{asiaee:2025}. Consequently, researchers frequently rely on observational data for CATE estimation, where causal effects can be identified only under additional assumptions such as no unobserved confounding. 

In time-to-event settings, treatment effect estimation is further complicated by competing risks, where patients are at risk of failure due to several, mutually exclusive event types.  
Evaluating the effectiveness of a treatment in preventing the event of interest requires accounting for all competing events (i.e., risks), especially when patients have comorbidities \citep{curth:2023}. For instance, assessing a cancer treatment might require considering its effect on cardiovascular events \citep{young:2020:causal, curth:2023}. If the treatment increases cardiovascular mortality, it automatically results in fewer deaths due to cancer, the cause of interest, even if the treatment has no benefit for cancer outcomes. Consequently, neglecting competing risks and judging the treatment as effective based solely on a lower incidence of cancer-related deaths would be misleading. 
 Treating competing events as censoring leads to an overestimation of the occurrence rate of the event of interest \citep{schuster:2020:ignoring, shen:2024:assessing}. \citet{schuster:2020:ignoring} illustrate this problem based on data from the LASA study \citep{hoogendijk:2016:LASA, hoogendijk:2019:LASA}, where standard survival methods substantially increased the estimated risk of depression by incorrectly assuming that individuals who died remained at risk for future depression.
Despite these issues and their importance for correct analyses, competing risks are often neglected in medical research \citep{vanWalraven:2016:competing, schumacher:2016:competing}.

In this work, we focus on estimating CATEs from right-censored survival data with competing risks under binary treatment. We define the CATE as the covariate-conditional difference in the absolute risk for the event of interest at a fixed time. Since the CATE is a causal quantity and hence not directly observable, its estimation requires suitable modeling strategies. 
A flexible and increasingly popular class of such strategies are meta-learners, that reduce CATE estimation to a sequence of standard supervised learning tasks. Thereby, any supervised learning algorithm can be utilized.
However, the learners’ relative performances depend on the characteristics of the data, e.g. the size of the treatment groups \citep{salditt:2023}. As a result, several simulation studies have compared the effectiveness of meta-learners across different data settings \citep{kuenzel:2019:stx, knaus:2020:MLEofHCE, alberto:2022:ITEnonparam, xu:2023:handbook}. Despite this growing literature, a comprehensive, large-scale evaluation for time-to-event data with competing risks is still lacking. 

To address this gap, we conduct a simulation study comparing six meta-learners for the estimation of CATEs in the presence of competing risks: the S-, T-, X-, M-, R-, and DR-Learner \citep{kuenzel:2019:stx,kennedy:2023:drlearner,powers:2018:mlearner+, nie:2020:rlearner}. Depending on the learner, models for the absolute risk and/or the CATE are required. To this end, we study three approaches for risk modeling: Cox regression \citep{cox:1972:coxreg} with and without Lasso penalty \citep{tibshirani:1996:lasso}, and random survival forests for competing risks \citep{ishwaran:2014:rsfcr}. Moreover, for direct CATE-modeling, we consider elastic net regression \citep{zou:2005:elnet} and random forests \citep{breiman:2001:rf}. 
The simulation design is inspired by the work of \citet{xu:2023:handbook}, who evaluated different meta-learners in classical survival settings without competing risks. We systematically vary hazard complexity, treatment heterogeneity, treatment assignment, event-type distribution and censoring mechanism. Moreover, we compare our findings to the results obtained by \citet{xu:2023:handbook} in the non-competing risks case. Our findings provide guidance on the choice of suitable meta-learners for causal inference in competing-risks settings. 

To facilitate applied use, this paper is further accompanied by a novel \texttt{R}-package \texttt{crsurvlearners}\footnote{\url{https://github.com/d-klippert/crsurvlearners}}. It implements the discussed methods, enables reproduction of the simulation results, and supports flexible generation of synthetic competing risks data under varying event-type ratios, propensity score specifications, censoring rates and covariate dimensionalities. Additionally, we demonstrate its 
practical applicability 
using a real-world dataset on Hodgkin's disease \citep{ishwaran:2023:rfsrc}. We conduct a subgroup analysis by fitting a regression tree to the predicted CATEs. This ``fit-the-fit'' approach, similar to that used by \citet{Logan:2019:ITEbarts, hu:2021:estimating, hu:2021:lungcancer}, allows us to examine average treatment effects within subgroups.

This paper is based on and extends the master's thesis \cite{klippert:2024} of the first author, which was supervised by the other two co-authors. The remainder of this paper is structured as follows.
In \Cref{sec:2}, we formally define the CATE framework for competing risks and outline necessary assumptions.
\Cref{sec:3} introduces the considered meta-learners.
We describe the simulation study in \Cref{sec:4} and discuss the corresponding results in \Cref{sec:5}.
In \Cref{sec:6}, we illustrate practical CATE estimation using a real-world dataset while providing code examples.
Finally, \Cref{sec:7} concludes the paper with a summary of our presented work.

\subsection{Related Work}
Heterogeneous treatment effects have gained increasing attention in recent years due to the growing demand for personalized treatment decision-making. For their estimation, a range of approaches has been proposed. Prominent examples include tree-based approaches such as honest regression trees \citep{athey:2016:recursive}, causal forests \citep{wager:2018:estimation}, Bayesian additive regression trees \citep{chipman:2010:bart}, Bayesian causal forests \citep{hahn:2020:bayesian}, causal survival forests \cite{cui:2023;estimating}, and causal boosting \citep{powers:2018:mlearner+}. In addition, \citet{powers:2018:mlearner+} considered treatment effect estimation based on multivariate adaptive regression splines.

These methods, however, are restricted in the choice of the underlying supervised learning algorithms. In contrast, a prominent class of methods that is model-agnostic are so-called meta-learners, such as the S-, T-, X-, M-, R-, and DR-Learners \citep{kuenzel:2019:stx, kennedy:2023:drlearner, powers:2018:mlearner+, nie:2020:rlearner}, which reduce the CATE estimation problem to a sequence of general supervised learning tasks. Several extensions of these meta-learners have been proposed to improve robustness and flexibility. For instance, \citet{li:2021:robust} reformulate the CATE estimation problem via general estimating equations, enabling robust estimation of heterogeneous treatment effects, including improved robustness to outliers for the M-Learner and adaptation to high-dimensional settings for the R-Learner. This work is extended by \citet{li:2023:robust} to additionally address model misspecification.

Since causal effects at the individual level cannot be observed in practice, the evaluation of HTE models is not feasible. This has led to a vast amount of simulation studies, methodological guides and reviews to assist practitioners in model selection \cite{powers:2018:mlearner+,wendling:2018:comparing,carvalho:2019:assessing,knaus:2020:MLEofHCE,curth:2021:nonparametric,jacob:2021:meetsML,zhang:2021:unified,alberto:2022:ITEnonparam,xu:2023:handbook,bo:2025:evaluating,moller:2026:estimating}. In the context of time-to-event data, \citet{xu:2023:handbook} extended general-cause meta-learners to the right-censored survival context and provide a comprehensive comparison across a range of standard survival settings. However, a systematic evaluation in the presence of competing risks is still lacking, which we address with this work.

The problem of estimating treatment effects in the presence of competing risks has primarily been studied in the context of average treatment effects \citep{young:2020:causal, ozenne:2020:estimation, cantagallo:2021:new}. \citet{li:2024:doubly} extend this line of work to heterogeneous effects by proposing a doubly robust approach based on cumulative incidence functions and targeted maximum likelihood estimation. Moreover, \citet{frauen:2025:orthogonal} have recently developed a model-agnostic approach for estimating HTEs in survival settings. 

In this work, we focus on HTE estimation under the assumption of no unobserved confounding. However, methods have been proposed to address violations thereof, for instance by leveraging proxy variables \cite{sverdrup:2023:proximal} or instrumental variables \cite{kallus:2019:interval,syrgkanis:2019:instruments,frauen:2023:estimating}. Specifically tailored to time-to-event settings, \citet{pmlr:meir25} recently introduced an approach based on recursively imputed survival trees which extends to the instrumental variable setting.

\subsection{Contributions} 
The main contributions of this work are as follows:
\begin{itemize}
    \item We present the first systematic comparison of meta-learners for CATE estimation in survival settings with competing risks.
    \item We assess the performance of six meta-learners across multiple scenarios covering varying hazard complexity, treatment effect heterogeneity, treatment assignment, event-type distributions and censoring mechanism. This way we provide practical guidance for method selection.
    \item  We introduce the novel \texttt{R} package \texttt{crsurvlearners}, which implements all considered methods and supports reproducibility and applied analyses. We illustrate its applicability on a real-world dataset on Hodgkin's disease.
\end{itemize}

\section{The CATE Model for Competing Risks} \label{sec:2}
We investigate CATE estimation from right-censored survival times with competing risks under a binary treatment. We briefly recall the standard competing risks framework in \Cref{sec:2.1}, before introducing our CATE formulation including a motivational example in \Cref{sec:2.2} and discussing necessary assumptions in \Cref{sec:2.3}. 
Throughout, random variables are denoted with capital letters and their realizations with lowercase letters. Moreover, we use bold letters to indicate (random) vectors.

\subsection{Survival Analysis with Competing Risks} \label{sec:2.1}
\begin{figure}[ht]
\centering
\begin{tikzpicture}[node distance=0.8cm, >=stealth]
    \node (0) [draw, rectangle, font=\sffamily, align=center] {0 = Initial State};
    \node (1) [draw, rectangle, right=5cm of 0, yshift=2.4cm, font=\sffamily, align=center] {1 = Event 1 occured};
    \node (2) [draw, rectangle, below=of 1, font=\sffamily, align=center] {2 = Event 2 occured};
    \node (dots) [below=of 2, font=\sffamily, align=center] {$\vdots$};
    \node (J) [draw, rectangle, below=of dots, font=\sffamily, align=center] {$J$ = Event $J$ occured};

    \draw[->, thick] (0.east) -- node[above, font=\sffamily\small, pos=0.5, rotate=25] {$\alpha_{1}(t|\boldsymbol{z})$} (1.west);
    \draw[->, thick] (0.east) -- node[above, font=\sffamily\small, pos=0.5, rotate=12] {$\alpha_{2}(t|\boldsymbol{z})$} (2.west);
    \draw[->, thick] (0.east) -- node[above, font=\sffamily\small, pos=0.5, rotate=340] {$\alpha_{J}(t|\boldsymbol{z})$} (J.west);
\end{tikzpicture}
\caption{State dynamics of a competing risks process with $J$ competing events.}
\label{fig:2.1}
\end{figure}
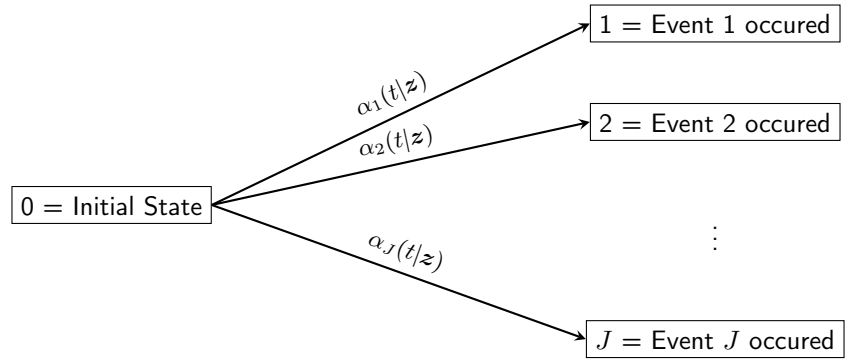

In a competing risks scenario, each individual is exposed to $J \in \mathbb{N}$ different event types, referred to as competing events. These can, for instance, reflect potential causes of death or failure. In contrast to traditional survival analysis, the focus lies not only on the time until an event occurs, but also on the specific type of the event. Following \citet{beyersmann:2013:handbook}, we model competing risks using a multistate model, as visualized in~\Cref{fig:2.1}. Therefore, we consider the competing risks process $(X_t)_{t \geq 0}$ with state space $\{0, \ldots, J\}$ and initial state 0, i.e., $P(X_0) = 1$. Then, the survival or event time $T = \inf\{t > 0 | X_t \neq 0\}$ is defined as the time until the first event $j \in \{1,\ldots, J\}$ occurs.

In a covariate-dependent setting with observed covariates $\boldsymbol{Z}=\boldsymbol{z} \in \mathbb{R}^p$, the stochastic dynamics of the competing risks process are characterized by the conditional cause-specific hazards \[\alpha_{j}(t|\boldsymbol{z})=\lim_{\Delta t \to 0} \dfrac{P(T < t+\Delta t, X_T = j | T \geq t, \boldsymbol{Z}=\boldsymbol{z})}{\Delta t}, \quad j \in \{1,\ldots,J\}.\] The overall or all-cause survival hazard is given by the sum of all cause-specific hazards \[\alpha(t|\boldsymbol{z}) = \sum_{j=1}^J \alpha_{j}(t|\boldsymbol{z})\]  and corresponds to the hazard function in the traditional single-event survival setting.
The concept of cumulative incidence functions (CIFs, also referred to as absolute risks) plays a central role in the estimation of CATEs introduced in~\Cref{sec:2.2}. For cause $j$, the CIF is the probability of experiencing event $j$ in the interval $[0,t]$. Given covariates $\boldsymbol{Z}=\boldsymbol{z}$, the CIF is defined as \begin{align} 
    F_j(t|\boldsymbol{z}) = P(T \leq t, X_T = j|\boldsymbol{Z}=\boldsymbol{z}) = \int^t_0 P(T > w-|\boldsymbol{Z}=\boldsymbol{z}) \alpha_{j}(w|\boldsymbol{z})dw,
\end{align} where $w-$ represents the left limit, i.e., $P(T > w-)=\lim_{s\uparrow w} P(T > s)$. Throughout this paper, we use the terms absolute risk and cumulative incidence interchangeably. 

In the presence of right-censoring, the observed event time is given by $U = \min\{T, C\}$, where $C$ denotes the censoring time.
Accordingly, we define the event indicator by \[ D = \begin{cases}
    X_T, & \text{if } U = T, \\
    0, & \text{if } U = C,
\end{cases} \] where $D=0$ implies censoring and $D=j$ denotes that event $j$ occurred.

\subsection{CATE Modeling - Setup and Motivation} \label{sec:2.2}
We compare different meta-learners for CATE estimation in a competing risks setting by means of a simulation study building upon the work of \citet{xu:2023:handbook}. Therefore, we consider settings with $n$ independent and identically distributed (i.i.d.) observations \[\{(\boldsymbol{Z}_i, A_i, U_i, D_i)\}_{i=1}^n.\] Here, $\boldsymbol{Z}_i \in \mathbb{R}^p$ denotes the vector of baseline covariates, $A_i \in \{0,1\}$ the binary treatment indicator, $U_i=\min \{T_i, C_i\}$ the observed (possibly right-censored) survival time, and $D_i$ the event indicator. 
The random variable $T_i$ corresponds to the survival time, while the random variable $C_i$ denotes the censoring time. 

Before formally defining the CATE, we begin with a motivating example to provide intuition for our set-up and the intended questions: Consider the evaluation 
of a newly developed medication ($A=1$) for the treatment of lung cancer compared to a standard treatment ($A=0$). Here, the event of interest is death due to lung cancer ($D=1$), while deaths from other causes, such as heart failure or diabetes, are competing ($D=2$). 
We want to investigate the causal effect of the new treatment on the survival of patients with specific characteristics $\boldsymbol{Z}=\boldsymbol{z}$, while accounting for other types of death. 
To formalize the causal question of this illustrative example, we assess the risk associated with the event of interest after a specific time span $t_0$, e.g. $t_0$ months after start of treatment. Let \[Y_i = \mathds{1}\{T_i \leq t_0, D_i=1\}\] represent the indicator for the event of interest at time $t_0$. In the above example, if the event of interest is defined as death due to lung cancer, $Y_i$ takes the value of 1 if individual $i$ experiences this outcome by time $t_0$ measured from the time origin, and 0 otherwise. 

Using the potential outcomes framework \citep{neyman:1923:framework, rubin:1974:framework}, we define the potential survival times of individual $i$ as $T_i(1)$ if given treatment 1 and $T_i(0)$ if given treatment 0. 
Analogously, let $D_i(a)$ represent the potential event type experienced by individual $i$ under treatment $a \in \{0,1\}$. We then define the potential outcomes $Y_i(a) = \mathds{1}\{T_i(a) \leq t_0, D_i(a)=1\}$, $a=0,1$, 
and the conditional average treatment effect (CATE) as \begin{align} \label{eq:2}
    \tau(\boldsymbol{z}) = \expval{Y_i(1)-Y_i(0) | \boldsymbol{Z}_i=\boldsymbol{z}}. 
\end{align} This CATE describes the expected difference of the absolute risks of the event of interest between the two treatment arms, conditional on covariates $\boldsymbol{z}$. 

Since the CATE $\tau(\boldsymbol{z})$ is defined in terms of unobservable potential outcomes, its estimation is not straightforward. Under identifiability assumptions (\Cref{sec:2.3}), however, it can be estimated.

\subsection{Assumptions and CATE Identification}  \label{sec:2.3}
In line with \citet{hill:2023:matching} and \citet{neal:2020:notes}, the identification of the conditional average treatment effects requires the following assumptions: \begin{itemize}
    \item \textbf{No Interference}:  $Y_i(a_1,\ldots, a_{i-1}, a_i, a_{i+1}, \ldots, a_n) = Y_i(a_i) $. \newline
    The outcome of an individual $i$ is unaffected by the treatment of other individuals $k \neq i$.
    \item \textbf{Consistency}: $\left( A=a \Rightarrow Y=Y(a) \right) \iff Y=Y(A)$. 
    \newline Multiple versions of treatments are precluded, i.e., treatment is uniquely defined.
    \item \textbf{Positivity}: $
        \forall \boldsymbol{z} : \text{ } 0 < P(A=a|\boldsymbol{Z}= \boldsymbol{z}) < 1 \text{, $a\in \{0,1\}$}$.
    \newline All subgroups have positive probability of receiving any treatment.
    \item \textbf{Conditional Exchangeability / Unconfoundedness}:
    $ (Y(0), Y(1)) \cindep A \, |\, \boldsymbol{Z} $.
    \newline All common causes, (confounders), of $A$ and $Y$ are measured by $\boldsymbol{Z}$. 
\end{itemize}

The assumptions of no interference and consistency are jointly known as SUTVA
(Stable Unit-Treatment Value Assumption).
The consistency assumption might seem trivial, however it might be violated if the treatment specification is too vague. For instance, in the lung cancer survival example, a patient might recover only if the drug dose of the new medication is sufficiently large. If the treatment $A=1$ only expresses that a patient receives the new medication, this might lead to different potential outcomes $Y(1)$ of the survival time depending on the drug dose. The preclusion of multiple versions of treatment thereby ensures that $Y(1)$ and $Y(0)$ are well-defined. 

Under these identifiability assumptions, we can rewrite the CATE as \begin{align}
\tau(\boldsymbol{z}) & = \expval{Y_i(1)-Y_i(0) | \boldsymbol{Z}_i=\boldsymbol{z}} \nonumber\\
& = \expval{Y_i|\boldsymbol{Z}_i=\boldsymbol{z}, A_i=1} - \expval{Y_i|\boldsymbol{Z}_i=\boldsymbol{z}, A_i=0} \nonumber \\
& = P(T \leq t_0, D_i=1|\boldsymbol{Z}_i=\boldsymbol{z}, A_i=1) - P(T \leq t_0, D_i=1|\boldsymbol{Z}_i=\boldsymbol{z}, A_i=0) \label{eq:2.3.1} \\
& = F_1(t_0 | \boldsymbol{Z}_i=\boldsymbol{z}, A_i=1) - F_1(t_0| \boldsymbol{Z}_i=\boldsymbol{z}, A_i=0), \nonumber
\end{align}
where $F_1(t|\boldsymbol{z},a)$ denotes the cause-specific CIF of the event of interest given covariates $\boldsymbol{z}$ and treatment $a \in \{0,1\}$. Hence, we can use methods for CIF estimation to directly estimate the CATE defined in \Cref{eq:2} which is the strategy behind the S-Learner (\Cref{sec:3.3}). To apply common aprroaches for CIF estimation, we additionally assume non-informative cenoring $C \cindep T \, |\, \boldsymbol{Z}, A$. 

As can be seen in \Cref{eq:2.3.1}, the CATE corresponds to the difference in the probabilities of experiencing the event of interest by time $t_0$ given covariates $\boldsymbol{z}$. Thus, the CATE can be utilized in treatment decision-making. If $\tau(\boldsymbol{z}) > 0$, the probability of dying by time $t_0$ due to the cause of interest, e.g., lung cancer, is smaller under treatment 0 than under treatment 1. As a result, treatment 0 should be recommended. Analogously, if $\tau(\boldsymbol{z}) < 0$, treatment 1 should be preferred. 

In randomized controlled trials, the unconfoundedness assumption is automatically satisfied, even if not all confounders are measured, because treatment assignment is randomized. In observational studies, by contrast, we rely on the assumption that all confounders are observed. We note, however, that there is a growing literature on estimating CATEs when this assumption is violated. Existing approaches include, for instance, methods based on instrumental variables \citep{syrgkanis:2019:instruments, frauen:2023:estimating} or proxies \citep{sverdrup:2023:proximal}. Another approach is to conduct a sensitivity analysis \citep{kallus:2019:interval}, providing bounds for the CATE function to assess robustness to unmeasured confounding.

\section{Meta-Learners} \label{sec:3}
The learning methods discussed in this section are referred to as meta-learners, as they estimate heterogeneous treatment effects using arbitrary supervised learning algorithms \citep{jacob:2021:meetsML}. This conversion of the causal effect estimation problem into a composition of well-understood supervised learning tasks, allows practitioners with domain knowledge to select suitable learning algorithms for the corresponding subproblems. For instance, many meta-learners require propensity score models which estimate the probability of receiving treatment $A=1$ given covariates. Such auxiliary functions are commonly referred to as nuisance components. In the competing risks setting considered here, nuisance components include absolute risk models, propensity score models, and inverse probability of censoring weight models. We discuss their estimation in \Cref{sec:3.2}, after introducing the necessary notation in \Cref{sec:3.1}, and subsequently present the individual meta-learning algorithms in Sections \ref{sec:3.3}--\ref{sec:3.8}.

\subsection{Notation} \label{sec:3.1}
To formalize the estimation of nuisance components and meta-learners, we introduce additional learning notation. We adopt the notation used by \citet{xu:2023:handbook} and write
\begin{align} \label{eq:3.1}
    \mathcal{M}\left( \widetilde{Y} \thicksim \widetilde{\boldsymbol{Z}}; \widetilde{\mathcal{O}}, [\widetilde{\boldsymbol{K}}] \right) 
\end{align} 
to represent a predictive model trained to model $\widetilde{Y}$ as a function of covariates $\widetilde{\boldsymbol{Z}}$. Here, $\widetilde{\mathcal{O}}$ denotes the training data, and $\widetilde{\boldsymbol{K}}$ corresponds to a vector of sample weights for all instances in the training data. By default, uniform weights are used. In this case, $\widetilde{\boldsymbol{K}}$ is not listed within the model representation in \Cref{eq:3.1}. Since this notation does not prescribe any specific model structure or modeling assumptions, it is well-suited for representing meta-learning models. Within the scope of this work, we focus on the S-, T-, X-, DR-, M- and R-Learner, as these represent widely used meta-learning strategies for CATE estimation. We present these meta-learning techniques specifically for the competing risks setting. Their practical use is illustrated in \Cref{sec:6}, where we show how to apply the discussed methods to a real-data example using the \texttt{crsurvlearners} package. 

Furthermore, we denote the available training data by \[ \mathcal{O} = \{(\boldsymbol{Z}_i, A_i,U_i, D_i) \, | \, i \in \mathcal{I} \}, \] where $\mathcal{I} = \{1,2,\ldots,n\}$. Additionally, we define the training datasets stratified by treatment-specific subsets as \begin{align*}
    \mathcal{O}_0 &= \{(\boldsymbol{Z}_i, A_i,U_i, D_i) \, | \, i \in \mathcal{I}: A_i = 0\} \text{ and} \\
    \mathcal{O}_1 &= \{(\boldsymbol{Z}_i, A_i,U_i, D_i) \, | \, i \in \mathcal{I}: A_i = 1\}.
\end{align*}

\subsection{Nuisance Components} \label{sec:3.2}

\subsubsection{Risk Models} \label{sec:3.2.1}
Most meta-learners depend on predictive models of the cumulative incidence function in survival settings. Using the notation introduced above, such a model can be expressed as \[ \widehat{\mu}(\cdot) = \mathcal{M}(Y \thicksim (\boldsymbol{Z}, A); \mathcal{O}), \] where $\widehat{\mu}(\boldsymbol{z}, a)$ denotes the predicted cause-1-specific cumulative incidence under treatment level $a$ for covariate vector $\boldsymbol{z}$. A range of modeling strategies can be used for its estimation.
Within this work, we focus on (penalized) Cox regression \cite{cox:1972:coxreg} and random survival forests for competing risks \cite{ishwaran:2014:rsfcr}.

\textbf{Cox regression} makes structural assumptions about the cause-specific hazard rate, which is then used to obtain the cumulative incidence function. Given baseline covariates $\boldsymbol{Z}_i$, the cause-$j$-specific hazard is modeled as 
\begin{equation} \label{eq:coxph}
    \alpha_{j}(t|\boldsymbol{Z}_i) = \alpha_{j; 0}(t)  \exp(\boldsymbol{\beta}_{j}^\top \boldsymbol{Z}_i),  
\end{equation}
where $\alpha_{j;0}(t)$ is an unspecified baseline hazard and $\boldsymbol{\beta}_{j} \in \mathbb{R}^p$ is a vector of unknown coefficients. The coefficients are estimated by maximizing the Cox partial likelihood. Besides this standard approach, we consider penalized estimation under the Lasso penalty \cite{tibshirani:1996:lasso}. The Cox Lasso regularizes the model by penalizing the absolute values of the coefficients, shrinking them towards zero and thereby acting as a variable selection mechanism.

In contrast, \textbf{random survival forests for competing risks} constitute a non-parametric approach. A total of $B$ trees are constructed, each grown on a bootstrap sample of the training data by recursively splitting the sample in two until any further partition would result in a node with less than $M$ individuals. In our simulation study, we utilize the composite split rule after Gray's test \citep{gray:1988:graytest}, which tests the null hypothesis \[ F_{j;l} (t) = F_{j;r} (t)\quad \forall \, t \leq u_{\text{max}},\] where $F_{j;l}$ and $F_{j;r}$ denote the covariate-independent cause-$j$-specific CIFs in the resulting daughter nodes and $u_{\text{max}}$ is the largest event time considered in the split. 
For each split point in each of the mtry randomly selected covariates, the split rule evaluates the heterogeneity across the cause-specific CIFs. The heterogeneity is quantified based on a weighted average of the Gray's test statistics over all causes, and the split which maximizes heterogeneity is selected. After construction, the cause-$j$-specific CIF for a new observation is obtained by averaging the Aalen-Johansen estimates \citep{aalen:1978:johansen-aalen} computed within the terminal node to which the observation is assigned across all trees.

\subsubsection{Propensity Score Models} 
Another nuisance component required by most meta-learners is the propensity score
\[ e(\boldsymbol{z}) = P(A=1|\boldsymbol{Z}=\boldsymbol{z}), \] which represents the probability of receiving treatment $A=1$ given covariates $\boldsymbol{z}$. The propensity score can be estimated using standard methods such as logistic regression, and its predictive performance can also be evaluated using established supervised learning diagnostics. Under the notation introduced above, the estimated propensity score based on the training data $\mathcal{O}$ can be expressed as the predictive model 
$
    \widehat{e}(\cdot) = \mathcal{M}\!\left( A \thicksim \boldsymbol{Z}; \mathcal{O} \right).
$

\subsubsection{Direct CATE Models and Inverse Probability of Censoring Weights}
To estimate the CATE, certain meta-learners such as the S-Learner and T-Learner combine nuisance components, here risk models, without directly modeling the CATE. Hence, these learning methods lack the capability to directly regularize the CATE. They only provide the opportunity to regularize the absolute risk, which can lead to a worse performance for the estimated CATE, as pointed out by \citet{xu:2022:survlearners}. Other meta-learning techniques including the X-, R-, and DR-Learner address this shortcoming by directly modeling the CATE in addition to the absolute risk enabling separate regularization. This is usually desirable, because it allows the CATE, which is typically smoother and sparser, to be regularized more aggressively than the absolute risk \citep{alberto:2022:ITEnonparam}. 

If we had access to both potential outcomes, the CATE could be learned directly by modeling the individual treatment effect (ITE) $ Y_i^* = Y_i(1) - Y_i(0)$ as a function of covariates resulting in the predictive model given by \[ \widehat{\tau}(\cdot) = \mathcal{M}(Y^* \thicksim \boldsymbol{Z}; \mathcal{O}). \] Even though $Y_i^*$ can never be observed (fundamental problem of causal inference), 
the aforementioned meta-learners are designed to bypass this ``missing data'' issue \citep{xu:2023:handbook} and involve direct models for the CATE. For estimating direct CATE models, we consider \textbf{random forests for regression} and \textbf{elastic net regression} \cite{zou:2005:elnet}, which applies regularization by combining the ridge \citep{hoerl:1970:ridge} and Lasso \citep{tibshirani:1996:lasso} penalties. 

However, another missing data problem arises from the fact that the indicator for the event of interest $Y_i = \mathds{1}\{T_i \leq t_0, D_i = 1\}$ is only observed for subjects which remain uncensored until time $t_0$. In line with \citet{xu:2023:handbook}, we refer to these subjects as complete cases and denote the corresponding index set and dataset by \begin{align*}
 \mathcal{I}_{\text{comp}} & = \{ i \in \mathcal{I} \, | \, C_i > \min \{t_0, T_i \} \} \text{ and} \\
 \mathcal{O}_{\text{comp}} & = \{ (\boldsymbol{Z}_i, A_i,U_i, D_i) \, | \,i \in \mathcal{I}_{\text{comp}}. \}  
\end{align*} Training a predictive model solely on the complete cases can induce selection bias in CATE estimation \citep{vock:2016:ipcw}. This problem is resolved through the use of inverse probability of censoring weights (IPCW), which adjusts the sample weights of complete observations based on estimated censoring probabilities. 

In order to estimate the censoring probability by time $t_0$, we train a predictive model for the censoring function $G(t|\boldsymbol{z}, a) = P(C > t|\boldsymbol{Z} = \boldsymbol{z}, A=a)$ based on all data points. We denote the resulting model by
\[ \widehat{G}(\cdot) = \mathcal{M}(\mathds{1}\{C \geq u\} \thicksim (\boldsymbol{Z}, A); \mathcal{O}). \] For instance, classical random survival forests \citep{ishwaran:2008:rsf} or, when censoring is independent of the covariates, the Kaplan-Meier estimate \citep{kaplan:1958:nonparametric} can be employed as the predictive model for the censoring function. Within the \texttt{crsurvlearners} package, both of these options are implemented. Each subject $i \in \mathcal{I}_{\text{comp}}$ is then assigned a weight which is calculated by \begin{align} \label{weight}
\widehat{K}_i = (\widehat{G}(\min\{U_i, t_0\}|\boldsymbol{Z_i}, A_i))^{-1}.
\end{align} These adjusted weights account for subjects with the same event time who were censored before time $t_0$ \citep{vock:2016:ipcw}. Subjects with smaller event times receive less weight, since they are less likely to be censored.

To avoid overfitting when using inverse probability of censoring weights, it is recommended to estimate these weights in an out-of-sample fashion to attain train and test set separation.
In order to calculate the out-of-sample weights, we first partition the training data into $k$ folds. Each fold is defined by one of the disjoint index sets $\mathcal{I}^{(1)}, \ldots, \mathcal{I}^{(k)}$. We then train $k$ models, with each model using $k-1$ folds as its training data.
For the $q$-th model ($1 \leq q \leq k$), all folds except $\mathcal{I}^{(q)}$ are used during training. 
The out-of-sample weight for the complete case $i_\ell \in \mathcal{I}^{(q)} \cap \mathcal{I}_\text{comp}$ is determined based on the model \[ \widehat{G}(\cdot) = \mathcal{M}(\mathds{1} \{C \geq u\} \thicksim (\boldsymbol{Z}, A); \mathcal{O}^{-q}), \] where $ \mathcal{O}^{-q} = \{(\boldsymbol{Z}_i, A_i,U_i, D_i) \, |\, i\in \mathcal{I}: i \notin \mathcal{I}^{(q)} \} $ corresponds to the data points which were not assigned to fold $q$. The weight $\widehat{K}_{\ell}$ is then calculated as in \Cref{weight}. 

\subsection{S-Learner} \label{sec:3.3}
Throughout this section, we describe the estimation of the CATE using the S-Learner \citep{hill:2011:bayesian, foster:2011:subgroup, kuenzel:2019:stx} which corresponds to direct estimation of the CATE through absolute risk modeling. First, the nuisance function \begin{align*}
    \mu(\boldsymbol{z}, a) = \expval{Y|\boldsymbol{Z}= \boldsymbol{z}, A=a}
\end{align*}
is estimated by a selected machine learning algorithm based on the training data $\mathcal{O}$. This yields the predictive model given by \[ \widehat{\mu}(\cdot) = \mathcal{M}(Y \thicksim (\boldsymbol{Z}, A); \mathcal{O}) . \] Then, the conditional average treatment effect $\tau(\boldsymbol{z}) = \mu(\boldsymbol{z}, 1) - \mu(\boldsymbol{z}, 0)$ is predicted by \begin{align*}
     \widehat{\tau}(\boldsymbol{z}) = \widehat{\mu}(\boldsymbol{z}, 1) - \widehat{\mu}(\boldsymbol{z}, 0).
\end{align*}
In the competing risks setting, $\mu(\boldsymbol{z}, a)$ is equal to the absolute risk given $\boldsymbol{z}$ and $a$, i.e., \[\mu(\boldsymbol{z}, a) = F_1(t_0|\boldsymbol{Z}_i=\boldsymbol{z}, A_i=a).\] Thus, methods for estimating cumulative incidence functions, e.g., Cox regression and random survival forests for competing risks are suitable choices as base-learners, as discussed in \Cref{sec:3.2.1}. 

The described meta-learner is referred to as S-Learner, since it involves only a single model fit. In this approach, the treatment variable is treated as any other covariate. Consequently, learning algorithms might disregard the treatment variable entirely. For example, methods based on random forests might select only variables different from the treatment variable when constructing new splits. This characteristic is advantageous when the true CATE is 0 for many $\boldsymbol{z}$, because ignoring the treatment variable results leads to $\widehat{\mu}(\boldsymbol{z}, 1) = \widehat{\mu}(\boldsymbol{z}, 0)$. 

\subsection{T-Learner}
Unlike the S-Learner, the T-Learner requires fitting two risk models. Hence its name, where ``T'' stands for ``two''. As stated by \citet{kuenzel:2019:stx} and \citet{xu:2023:handbook}, the method involves estimating the functions \begin{align} \label{mus} \mu_{(1)}(\boldsymbol{z}) = \expval{Y(1)|\boldsymbol{Z} = \boldsymbol{z}}, \quad \mu_{(0)}(\boldsymbol{z}) = \expval{Y(0)|\boldsymbol{Z} = \boldsymbol{z}} 
\end{align} separately by treatment, ensuring that the treatment variable is not ignored. This is done by training two predictive models \begin{align*}
\widehat{\mu}_{(1)}(\cdot) = \mathcal{M}(Y \thicksim \boldsymbol{Z}; \mathcal{O}_1), \quad \widehat{\mu}_{(0)}(\cdot) = \mathcal{M}(Y \thicksim \boldsymbol{Z}; \mathcal{O}_0),
\end{align*} where $\widehat{\mu}_{(1)}$ is trained based on the subgroup with treatment $A=1$, while $\widehat{\mu}_{(0)}$ is trained based on the individuals with treatment $A=0$. The CATE is then estimated as the difference in the predicted absolute risks given by \[ \tau(\boldsymbol{z})= \widehat{\mu}_{(1)}(\boldsymbol{z})- \widehat{\mu}_{(0)}(\boldsymbol{z}).\] 

A drawback of the T-Learner is its susceptibility to regularization bias \citep{kuenzel:2019:stx, nie:2020:rlearner}. When the CATE function is close to zero, the T-Learner must independently learn nearly identical nuisance functions $\mu_{(1)}$ and $\mu_{(0)}$ to capture a small treatment effect. As both nuisance functions are typically regularized separately, covariate effects present in both treatment groups may shrink differently across $\mu_{(1)}$ and $\mu_{(0)}$ which may lead to a biased CATE estimate.

\subsection{X-Learner}
The X-Learner \citep{kuenzel:2019:stx} estimates conditional average treatment effects based on the imputed individual treatment effects 
\begin{align*}
Y_i^{*,\boldsymbol{Z}, 1} &= Y_i(1) - \mu_{(0)}(\boldsymbol{Z}), \\
Y_i^{*,\boldsymbol{Z}, 0} &= \mu_{(1)}(\boldsymbol{Z}) - Y_i(0),
\end{align*}
where $\mu_{(a)}(\boldsymbol{z}) = \expval{Y_i(a)|\boldsymbol{Z}=\boldsymbol{z}}$ as in \Cref{mus} for $a=0,1$. Predictive models for these imputed individual treatment effects provide CATE estimates, since 
\begin{align*}
\expval{Y_i^{*,\boldsymbol{Z},a}|\boldsymbol{Z}_i = \boldsymbol{z}} = \expval{Y_i(1)-Y_i(0)|\boldsymbol{Z}_i = \boldsymbol{z}} = \tau(\boldsymbol{z}), \quad a=0,1.
\end{align*} Unlike the standard ITEs, imputed individual treatment effects can be ``observed'' for all complete cases by plugging in the estimates $\widehat{\mu}_{(1)}$ and $\widehat{\mu}_{(0)}$ for $\mu_{(1)}$ and $\mu_{(0)}$, respectively. We estimate $\widehat{\mu}_{(1)}$ and $\widehat{\mu}_{(0)}$ as in the T-Learner stratified by treatment, which corresponds to the first modeling step in the X-Learner. Thereby, we can access $Y_i^{*,\boldsymbol{Z}, 1}$ for individuals in treatment arm 1, whereas $Y_i^{*,\boldsymbol{Z}, 0}$ would be accessible in treatment arm 0.

In the second step, we train the predictive models
\begin{align*}
\widehat{\tau}_{(1)}(\cdot) & = \mathcal{M}(Y-\widehat{\mu}_{(0)}(\boldsymbol{Z}) \thicksim \boldsymbol{Z}; \mathcal{O}_1 \cap \mathcal{O}_{comp}, \widehat{\boldsymbol{K}} ) , \\
\widehat{\tau}_{(0)}(\cdot) & = \mathcal{M}(\widehat{\mu}_{(1)}(\boldsymbol{Z}) - Y \thicksim \boldsymbol{Z}; \mathcal{O}_0 \cap \mathcal{O}_{comp}, \widehat{\boldsymbol{K}} ).
\end{align*}
Here $\widehat{\boldsymbol{K}}$ are IPCWs which are necessary, since the training data has to be reduced to the complete cases. The model $\widehat{\tau}_{(a)}$ for $a\in \{0,1\}$ is trained solely on the complete cases from treatment group $a$, since the estimated imputed ITE $Y_i^{*,\boldsymbol{Z}, a}$ is not observable in the opposite treatment group.

In the final step, the resulting predictive models are combined to the weighted sum
\begin{align*}
\widehat{\tau}(\boldsymbol{z}) = (1-g(\boldsymbol{z})) \cdot \widehat{\tau}_{(1)} (\boldsymbol{z})+ g
(\boldsymbol{z}) \cdot \widehat{\tau}_{(0)}(\boldsymbol{z}).
\end{align*} The function $g$ with $g(\boldsymbol{z}) \in [0,1]$ for all $\boldsymbol{z}$ is a weighting function, which is chosen as the estimated propensity score $g(\boldsymbol{z}) = \widehat{e}(\boldsymbol{z})$ in this work. The intuition is to upweight $\widehat{\tau}_{(1)}$ when there are more observations with treatment $0$ and vice versa in order to improve prediction performance. When there are more cases in treatment group 0, $\widehat{\mu}_{(0)}(\boldsymbol{z})$ is estimated on a larger dataset than $\widehat{\mu}_{(1)}(\boldsymbol{z})$. Consequently, the imputed ITEs derived from $\widehat{\mu}_{(0)}(\boldsymbol{z})$ are expected to better capture the CATE. In this situation, the propensity score weighting ensures that $\widehat{\mu}_{(0)}(\boldsymbol{z})$ has greater impact on the final prediction than $\widehat{\mu}_{(1)}(\boldsymbol{z})$.

In direct comparison to the T-Learner, the X-Learner increases data efficiency \citep{neal:2020:notes}. The T-Learner only combines the information from both treatment groups by the difference between the estimates for $\mu_{(1)}$ and $\mu_{(0)}$ and therefore does not estimate a predictive model on the basis of the entire data. In contrast, the X-Learner combines the functions $\widehat{\tau}_{(1)}$ and $\widehat{\tau}_{(0)}$ which are estimated with information from both treatment groups. This has been shown to be advantageous when treatment assignment is unbalanced \citep{kuenzel:2019:stx}. In such settings, the X-Learner achieves a faster convergence rate than the T-Learner.

\subsection{M-Learner}
The M-Learner \citep{horvitz:1952:generalization, hirano:2003:efficient, knaus:2020:MLEofHCE, xu:2023:handbook} corresponds to the inverse probability weighting (IPW) estimator by \citet{horvitz:1952:generalization}. It is the only considered meta-learner that does not rely on a specification of a risk model. Instead, the CATE is directly estimated by modeling the modified outcomes \begin{align*}
Y_i^{*,M} = Y_i \left(\dfrac{A_i}{e(\boldsymbol{Z_i})} - \dfrac{1-A_i}{1-e(\boldsymbol{Z_i})}\right), \quad i \in \mathcal{O}_{\text{comp}}. 
\end{align*} These modified outcomes give the M-Learner its name. If the propensity score is unknown as typically the case with observational data, it needs to be estimated in order to construct the modified outcomes. Similar to the IPCWs, out-of-sample estimates for the propensity score are required to avoid overfitting \citep{jacob:2021:meetsML}. Since the modified outcomes can only be constructed for complete observations, again, the use of inverse probability of censoring weights $\widehat{\boldsymbol{K}}$ is required. The predictive model for the CATE is specified by \begin{align} \label{eq:3.2}
\widehat{\tau}(\cdot) = \mathcal{M}(Y^{*,M} \thicksim \boldsymbol{Z}; \mathcal{O}_{\text{comp}}, \widehat{\boldsymbol{K}}).
\end{align}

Due to the equality $\expval{Y_i^{*,M}|\boldsymbol{Z}=\boldsymbol{z}} = \tau(\boldsymbol{z})$ shown in \citet{powers:2018:mlearner+}, the M-Learner allows for unbiased CATE estimation. However, it suffers from high variance, because the modified outcomes blow up for propensity score estimates close to 0 or 1. In comparison, the X-Learner and R-Learner (\Cref{sec:3.7}) can estimate the CATE with lower variance \citep{xu:2023:handbook}, but require models of the absolute risk.  

\textbf{Remark} It is apparent from the definition of the modified outcomes in \Cref{eq:3.2} that $Y_i^{*,M} \in (-\infty, -1) \cup \{0\} \cup (1,\infty)$. 
It follows that if $Y_i = 0$, then $Y_i^{*,M}=0$. However, when $Y_i=1$, the modified outcome becomes larger than 1 in absolute value, because the propensity score satisfies $0 < e(\boldsymbol{Z}_i) < 1$. Consequently, the M-Learner models the CATE based on modified outcomes that are either outside of this range or equal 0, although the CATE is limited to the range $[-1,1]$ due to it being the difference of two probabilities. Therefore, practitioners should be cautious when using the M-learner in the considered competing risks survival setting, as the modified outcome construction can lead to CATE estimates outside the admissible range $[-1,1]$. 

\subsection{R-Learner} \label{sec:3.7}
The R-Learner \citep{nie:2020:rlearner} estimates CATEs by minimizing a loss function built upon covariate-based residuals for $A$ and $Y$. The residualization originates from \citet{robinson:1988:rootNconsistent} within the context of partially linear models and was later adopted by \citet{chernozhukov:2018:double} within average treatment effect estimation. The residuals for $A_i$ and $Y_i$ are obtained by subtracting their conditional means given $\boldsymbol{Z}_i$, denoted by the propensity score $e(\boldsymbol{Z}_i)$ and $\expval{Y_i | \boldsymbol{Z}_i}$.
For the latter, we define the quantity
 \begin{align*}
m(\boldsymbol{z}) & =  \expval{Y_i | \boldsymbol{Z}_i = \boldsymbol{z}} \\
&= \expval{Y_i|\boldsymbol{Z}_i=\boldsymbol{z}, A_i=1} P(A_i=1|\boldsymbol{Z}_i=\boldsymbol{z}) + \expval{Y_i|\boldsymbol{Z}_i=\boldsymbol{z}, A_i=0} P(A_i = 0|\boldsymbol{Z}_i=\boldsymbol{z}) \\
&= e(\boldsymbol{z}) \mu_{(1)}(\boldsymbol{z}) + (1-e(\boldsymbol{z})) \mu_{(0)}(\boldsymbol{z}),
\end{align*} which allows us to write the residuals as $Y_i - m(\boldsymbol{Z}_i)$ and $A_i - e(\boldsymbol{Z}_i)$. The R-Learner received its name from \citet{nie:2020:rlearner} in recognition of the original contribution by \citet{robinson:1988:rootNconsistent} and the importance of the residuals. The two residuals are linked by the expectation of $Y_i - m(\boldsymbol{Z}_i)$ conditioned on $\boldsymbol{Z}_i = \boldsymbol{z}$ and $A_i=a$, given by
\begin{align*}
& \, \expval{Y_i-m(\boldsymbol{Z}_i)|\boldsymbol{Z}_i=\boldsymbol{z}, A_i=a} \\  
= & \,  \expval{Y_i|\boldsymbol{Z}_i = \boldsymbol{z}, A_i=a} - e(\boldsymbol{z}) \mu_{(1)}(\boldsymbol{z}) - (1-e(\boldsymbol{z})) \mu_{(0)}(\boldsymbol{z})\\
= &\,  \mu_{(a)}(\boldsymbol{z}) - \mu_{(0)}(\boldsymbol{z}) - e(\boldsymbol{z}) \tau(\boldsymbol{z}) \\
= &\,  a \mu_{(1)}(\boldsymbol{z}) + (1-a)  \mu_{(0)}(\boldsymbol{z}) - \mu_{(0)}(\boldsymbol{z}) - e(\boldsymbol{z}) \tau(\boldsymbol{z}) \\
= &\,  (a-e(\boldsymbol{z}))\tau(\boldsymbol{z}).
\end{align*} This relationship motivates the loss-based representation of the CATE, given by 
\begin{align}
\tau(\cdot) & \in \text{argmin}_{\tilde{\tau}(\cdot)} \, \expval{\left((Y_i-m(\boldsymbol{Z}_i))-(A_i-e(\boldsymbol{Z}_i))\tilde{\tau}(\boldsymbol{Z}_i)\right)^2}  \label{LOSSBASED} \\ 
& = \text{argmin}_{\tilde{\tau}(\cdot)} \, \expval{(A_i-e(\boldsymbol{Z}_i))^2\left(\dfrac{Y_i-m(X_i)}{A_i-e(\boldsymbol{Z}_i)}-\tilde{\tau}(\boldsymbol{Z}_i)\right)^2}. \nonumber
\end{align} 
Hence, the CATE can be estimated through a predictive model relating the covariates $\boldsymbol{Z}_i$ to the ratio of the residuals $Y_i - m(\boldsymbol{Z}_i)$ and $A_i - e(\boldsymbol{Z}_i)$. In order to compute this ratio, estimates of $\mu_{(1)} $, $\mu_{(0)}$ and $e$ (if unknown) are needed. To prevent overfitting, these functions are estimated in an out-of-sample manner. Aside from the out-of-sample estimation, $\mu_{(1)} $ and $\mu_{(0)}$ are estimated as in the T-Learner.

Following \citet{xu:2023:handbook}, we denote the target quantity in the R-Learner by \begin{align*}
Y^{*,R}_i = \dfrac{Y_i-\widehat{m}(\boldsymbol{Z}_i)}{A_i-\widehat{e}(\boldsymbol{Z}_i)}, \quad i\in \mathcal{I}_{\text{comp}}.
\end{align*} Since $Y^{*,R}_i$ can only be constructed for complete observations, IPCWs are required to mitigate censoring-induced selection bias. The expectation in the second line of \Cref{LOSSBASED} demonstrates that the inverse probability of censoring weights need to be further adjusted. To match the loss-based representation, the inverse probability of censoring weight $\widehat{K}_i$ for observation $i \in \mathcal{I}_{\text{comp}}$ has to be multiplied by $(A_i-\widehat{e}(\boldsymbol{Z}_i))^2$. We denote the resulting weights by $\widehat{\boldsymbol{K}}^{R}$.
The CATE is then estimated through the predictive model
\begin{align*}
\widehat{\tau}(\cdot) = \mathcal{M}(Y^{*,R} \thicksim \boldsymbol{Z}; \mathcal{O}_\text{comp}, \widehat{\boldsymbol{K}}^{R}).
\end{align*} 
Moreover, the R-Learner can perform well even if $\widehat{m}(\cdot)$ is not an accurate approximation of $m(\cdot)$ \citep{xu:2023:handbook}. To be more precise, if $\widehat{m}(\cdot) \approx 0$ and $\widehat{e}(\cdot) = 0.5$, the R-Learner's predictions are very similar to the predictions of the M-Learner. However, when $\widehat{m}(\cdot) \approx m(\cdot)$, the centering of $Y$ around $\widehat{m}(\cdot)$ enables the R-Learner to estimate the CATE with lower variance compared to the M-Learner.

\subsection{DR-Learner} \label{sec:3.8}
The DR-Learner \citep{kennedy:2023:drlearner} estimates the CATE via a pseudo-outcome regression that integrates both the propensity score and treatment-specific risk models. The pseudo-outcomes combine the T-Learner prediction $\widehat{\mu}_{(1)}(\boldsymbol{Z}_i) - \widehat{\mu}_{(0)}(\boldsymbol{Z}_i)$ with an IPW estimate on the residual $Y_i - \widehat{\mu}_{(A_i)}(\boldsymbol{Z}_i)$. The resulting estimator is doubly robust in the sense that accurate CATE estimation is achieved if either nuisance component is correctly specified. To be more specific, the DR-Learner is an unbiased estimator of the CATE if $\widehat{e}(\boldsymbol{z}) =e(\boldsymbol{z})$ or $\widehat{\mu}_{a}(\boldsymbol{z}) =\mu_{a}(\boldsymbol{z})$ for $a=0,1$. 

Prior to any model fitting, the training data is divided into two equally large folds which we identify by the index sets $\mathcal{I}^{(1)}$ and $\mathcal{I}^{(2)}$. We denote the corresponding datasets by $\mathcal{O}^{(1)}$ and $\mathcal{O}^{(2)}$ and describe the CATE estimation through a four-step algorithm consisting of the following steps:
\begin{itemize}
\item[1)] Estimate a model for the propensity score based on $\mathcal{O}^{(1)}$:
 \[ \widehat{e}(\cdot) = \mathcal{M}(A \thicksim \boldsymbol{Z}; \mathcal{O}^{(1)}). \] 
\item[2)] Fit models for $\mu_{(1)}$ and $\mu_{(0)}$ based on $\mathcal{O}^{(1)}$: 
\begin{align*}
 \widehat{\mu}_{(1)}(\cdot) & = \mathcal{M}(Y\thicksim \boldsymbol{Z}; \mathcal{O}_1 \cap \mathcal{O}^{(1)}) , \\
 \widehat{\mu}_{(0)}(\cdot) & = \mathcal{M}(Y\thicksim \boldsymbol{Z}; \mathcal{O}_0 \cap \mathcal{O}^{(1)}).
\end{align*}
\item[3)] For the complete cases $i$ from fold $\mathcal{I}^{(2)}$, i.e., $i \in \mathcal{I}_\text{comp} \cap \mathcal{I}^{(2)}$ construct the pseudo-outcome \begin{align*}
\widehat{\varphi}_i = \dfrac{A_i-\widehat{e}(\boldsymbol{Z}_i)}{\widehat{e}(\boldsymbol{Z}_i)(1-\widehat{e}(\boldsymbol{Z}_i))} \left( Y_i - \widehat{\mu}_{(A_i)}(\boldsymbol{Z}_i) \right) + \widehat{\mu}_{(1)}(\boldsymbol{Z}_i) - \widehat{\mu}_{(0)}(\boldsymbol{Z}_i).
\end{align*} Then, train a predictive model with inverse probability of censoring weights $\widehat{\boldsymbol{K}}$, again estimated in an out-of-sample manner, to relate the pseudo-outcomes to covariates:
\begin{align*}
\widehat{\tau}(\cdot) = \mathcal{M}(\widehat{\varphi} \thicksim \boldsymbol{Z} ;\mathcal{O}_\text{comp} \cap \mathcal{O}^{(2)}, \widehat{\boldsymbol{K}}).
\end{align*}
\item[4)] Repeat steps 1--3 by swapping the roles of the folds $\mathcal{I}^{(1)}$ and $\mathcal{I}^{(2)}$, i.e., use dataset $\mathcal{O}^{(2)}$ to estimate $e$, $\mu_{(1)}$ and $\mu_{(0)}$ and $\mathcal{O}^{(1)}$ for modeling the pseudo-outcomes. 
\end{itemize}
Let $\widehat{\tau}^{(1)}$ and $\widehat{\tau}^{(2)}$ represent the resulting predictive models. The final CATE estimate is computed as the mean prediction over these models, given by \[ \widehat{\tau}(\boldsymbol{z}) = \dfrac{1}{2} \left(\widehat{\tau}_{(1)} (\boldsymbol{z}) + \widehat{\tau}_{(2)} (\boldsymbol{z})\right). \] The fourth step, which performs cross-fitting, can be considered optional. Since cross-fitting increases data efficiency and yields more stable estimates, we include this step in our implementation within the \texttt{crsurvlearners} package.

\section{Simulation Design} \label{sec:4}
In this section, we describe the design of our simulation study. We compare the predictive performance of the S-, T-, X-, M-, R- and DR-Learner across a wide range of settings that vary in hazard complexity, treatment heterogeneity, treatment assignment, event-type distribution and censoring mechanism. The data generating process described in \Cref{sec:4.1} is inspired by the simulation framework of \citet{xu:2023:handbook}, who study meta-learners for CATE estimation in a standard survival setting without competing risks. We extend their framework to competing risks data and additionally include the DR-Learner in the comparison.

For absolute risk estimation within each meta-learner, we consider Cox regression with and without the Lasso penalty, as well as random survival forests (RSFs) for competing risks. For direct CATE modeling, we employ random forests and elastic net regression. Combining each absolute risk model with each CATE model for every applicable meta-learner yields a total of 26 distinct meta-learning algorithms. These algorithms are summarized in \Cref{Tab:Learners}, where it is important to note that not all model combinations are applicable to every meta-learner, since the S- and T-Learner do not model the CATE directly and the M-Learner does not involve a risk model. 

Following \citet{xu:2022:survlearners}, we denote each meta-learning algorithm using a shorthand notation consisting of three letters. The first letter indicates the meta-learner with the DR-Learner represented by  ``D''. The second letter denotes the absolute risk model, with ``C'' indicating Cox regression, ``C*'' indicating Cox regression with Lasso penalty, and ``S'' indicating random survival forests for competing risks. The third letter describes the CATE model, with ``E'' representing elastic net regression and ``R'' random forests. The symbol ``--'' indicates that the meta-learning algorithm does not incorporate a model of the respective type. For instance, ``SC--'' represents the S-Learner using Cox regression for absolute risk modeling, which does not include a direct model for the CATE. This notation facilitates reporting of results and is summarized in \cref{Tab:Learners}.
\begin{table}[ht]
\centering
\caption{Meta-learning algorithms considered in the simulation study. Each algorithm is defined by a combination of meta-learner, risk model and CATE model.}
\label{Tab:Learners}
\begin{tabular}{l|l|l|l}
Risk Model \textbackslash \, CATE Model& Random Forest (R) & Elastic Net (E) & None (--) \\ \hline
 Cox Regression (C) & XCR, RCR, DCR & XCE, RCE, DCE & SC--, TC-- \\ \hline
 Cox Lasso (C*) & XC*R, RC*R, DC*R & XC*E, RC*E, DC*E & SC*--, TC*--\\ \hline
 RSF for Competing Risks (S) & XSR, RSR, DSR & XSE, RSE, DSE & SS--, TS-- \\ \hline
 None (--) & M--R & M--E & \multicolumn{1}{c}{/}\\
\end{tabular}
\end{table}

For the meta-learners that require IPCWs (X-, M-, R-, DR-Learner), we estimate them using out-of-sample Kaplan-Meier estimators based on a 10-fold partition of the training data. Similarly, we estimate the propensity score within the M- and R-Learners using logistic regression in a 10-fold out-of-sample manner. Details on hyperparameter tuning for all supervised learning tasks are provided in Appendix \ref{sec:A.hyperparams}, whereas the evaluation metrics are defined in \Cref{sec:4.2}.

\subsection{Data Generating Process} \label{sec:4.1} 
We examine the performance of the meta-learning algorithms across 33 different data generating processes (DGPs). The different processes result from varying certain parameters of a base simulation setting while keeping others fixed. We first describe the general DGP and the parametrization of the base case. Subsequently, we cover the remaining configurations in Sections \ref{sec:4.1.1}--\ref{sec:4.1.4}. An overview of all parameter configurations is provided in \Cref{A:Tabelle} in Appendix \ref{sec:A.configs}.

For each configuration, we generate independent and identically distributed data of the form 
\begin{align*}
\{ (\boldsymbol{Z}_i, A_i, U_i, D_i) \}_{i=1}^n,
\end{align*} 
where $n = 10,000$. Half of the observations are used for training and the remaining half for testing. For complete cases, we set $Y_i = \mathds{1}\{T_i \leq t_0, D_i =1 \}$ with $t_0=1.2$. The time horizon of interest was chosen such that approx. two thirds of the sample have experienced an event by $t_0$ in the base setting.   

The simulation settings are inspired by \citet{xu:2023:handbook} and the corresponding R package \texttt{survlearners} \citep{xu:2022:survlearners}. In the first step of the DGP, covariates $\boldsymbol{Z}_i = (Z_{i1},\ldots,Z_{ip})^\top \in \mathbb{R}^p$ are drawn from a standard multivariate normal distribution, i.e., $Z_{ij} \overset{\text{i.i.d.}}{\thicksim} \mathcal{N}(0,1)$ for $j=1,\ldots,p$, with $p=25$ as in \citet{xu:2023:handbook}. Based on the covariates $\boldsymbol{Z}_i$, the treatment assignment is generated via $A_i \thicksim \text{Bernoulli}(e(\boldsymbol{Z}_i))$. In the base cases we set $ e(\boldsymbol{Z}_i) = 0.5$, i.e., the treatment assignment is balanced.

To simulate survival times, we utilize the hazard rate used in \citet{xu:2022:survlearners} as the all-cause hazard $\alpha_{0\cdot}$. Given covariates $\boldsymbol{Z}_i$ and treatment $A_i$, it is specified as
\begin{align} \label{eq:all-cause}
\alpha(t|\boldsymbol{Z}_i,A_i) = \dfrac{1}{2\sqrt{t}} \exp(f_R(\boldsymbol{Z}_i) + f_\tau(\boldsymbol{Z}_i, A_i)).
\end{align}
In accordance with \citet{xu:2023:handbook}, we refer to $f_R$ as the baseline risk function and to $f_\tau$ as the treatment covariate interaction function. If both functions are linear combinations of the covariates (and the treatment), the all-cause hazard corresponds to the hazard function of a Cox-Weibull model \citep{bender:2005:generating} with scale $a=1$ and shape $\nu = 0.5$, which decreases over time. In the base case, $f_R$ and $f_\tau$ are defined as \begin{align*}  
f_R(\boldsymbol{Z}_i) & =  \sum\limits_{j=1}^{25} \beta_1 \dfrac{Z_{ij}}{\sqrt{p}}  \\ f_\tau(\boldsymbol{Z}_i, A_i) &= (-0.5-\gamma_1 Z_{i2})A_i
\end{align*} 
with $\beta_1=1$ and $\gamma_1 = 0.5$. 

We consider two competing events in our simulations and set the cause-specific hazards as 
\begin{align*}
\alpha_{1}(t|\boldsymbol{Z}_i,A_i)& = \theta \cdot \alpha(t|\boldsymbol{Z}_i,A_i) 
\text{ and} \\
\alpha_{2}(t|\boldsymbol{Z}_i,A_i)& = (1-\theta) \cdot \alpha(t|\boldsymbol{Z}_i,A_i),
\end{align*} 
where $\theta \in [0,1]$ controls the proportion of event types.
In the base case, we set $\theta = 0.5$ resulting in approximately the same number of type 1 and type 2 events.

To simulate survival data in the presence of competing risks, we use the algorithm described in \citet{beyersmann:2009:simulating}. First, the survival time is generated from the Cox-Weibull model of the all-cause hazard.
Due to 
\begin{align} \label{proportion}
P(X_{T_i} = 1 | T_i, \boldsymbol{Z}_i, A_i) = \dfrac{\alpha_{1}(T_i|\boldsymbol{Z}_i, A_i)}{\alpha(T_i|\boldsymbol{Z}_i, A_i)} = \theta,
\end{align} 
we then run a Bernoulli experiment to generate the tentative event $X_{T_i}$, where a type 1 event occurs with probability $\theta$.

To prevent very large survival times and ensure numerical stability, we truncate the survival time via 
\[ 
T_i = \min\{12+E_i, \widetilde{T}_i\}, 
\] 
where $E_i \sim \mathcal{N}(0, 0.001^2)$ is added to prevent ties.

Censoring times are generated independently from a Weibull distribution, as defined in \citet{bender:2005:generating}, with constant hazard function. The base case parametrization is characterized by shape parameter $\nu=1$ and scale parameter $\lambda=0.1$ resulting in about $20\%$ censored observations. Then, the right-censored survival time is given by $U_i = \min\{T_i, C_i\}$, and the observed event type $D_i$ is determined by \begin{align*} D_i = \begin{cases}
    X_{T_i}, & \text{if } U_i = T_i, \\
    0, & \text{if } U_i = C_i.
\end{cases}
\end{align*} 

Finally, the true underlying CATE for each test data point is approximated via Monte Carlo: For an individual with covariates $\boldsymbol{Z}_i$, the true CATE is estimated based on 10,000 generated survival times with equal numbers assigned to each treatment group. 
To enable CATE estimation, we further assume the identifiability conditions No Interference, Consistency, Positivity and Unconfoundedness as described in \Cref{sec:2.3}. The validity of all these assumptions follows directly from the employed DGPs.

\subsubsection{Varying Complexity of the Hazard Functions} \label{sec:4.1.1}
The complexity of the hazard functions is adjusted by changing the definitions of $f_R$ and $f_\tau$ in the all-cause hazard defined in \Cref{eq:all-cause}. We differentiate between the function type, linear and nonlinear, as well as the number of covariates 
entering the functions $f_R$ and $f_\tau$, with $p_R, p_\tau \in\{1, 25\}$. 

The variations of the baseline risk function $f_R$ are given by: 
\begin{align}
\text{Linear, } p_R=1 &:   f_R(\boldsymbol{Z}_i) = \beta_1 Z_{i1}, \nonumber \\
\text{Linear, } p_R=25&: f_R(\boldsymbol{Z}_i) =  \sum\limits_{j=1}^{25} \beta_1 \dfrac{Z_{ij}}{\sqrt{p}}, \nonumber \\
\text{Nonlinear, } p_R=1 &:  f_R(\boldsymbol{Z}_i) = \beta_2\mathds{1}\{Z_{i1} > 0.5\}, \nonumber \\
\text{Nonlinear, } p_R=25&: f_R(\boldsymbol{Z}_i)= \beta_2 \mathds{1}\{ Z_{i1} > 0.5\} \nonumber \\ 
&  \phantom{: f_R(\boldsymbol{Z}_i)=} + \sum\limits_{j=1}^{12} \beta_3 \mathds{1} \{ Z_{i(2j)}>0.5\} \mathds{1} \{ Z_{i(2j+1)}>0.5\}. \nonumber
\end{align}
Here, we set $p=25$, $\beta_1 = 1$, $\beta_2=0.99$ and $\beta_3=0.33$. 

The treatment-covariate interaction functions $f_\tau$ are defined by:
\begin{align*}
\text{Linear, } p_\tau=1 &:   f_\tau(\boldsymbol{Z}_i, A_i) = (-0.5-\gamma_1 Z_{i2})A_i, \\
\text{Linear, } p_\tau=25&: f_\tau(\boldsymbol{Z}_i, A_i) =  \left(-0.5-\sum\limits_{j=1}^{25}  \gamma_1 \dfrac{Z_{ij}}{\sqrt{p}}\right)A_i, \\
\text{Nonlinear, } p_\tau=1 &:  f_\tau(\boldsymbol{Z}_i, A_i) =\left(-0.5 - \gamma_1 \mathds{1}\{Z_{i1} > 0.5\}\right) A_i, \\
\text{Nonlinear, } p_\tau=25&: f_\tau(\boldsymbol{Z}_i, A_i)= \Big(-0.5 - \gamma_2 \mathds{1}\{Z_{i1} > 0.5\} \\ 
&  \phantom{:f_\tau(\boldsymbol{Z}_i, A_i)=} - \sum\limits_{j=1}^{12} \gamma_3 \mathds{1} \{ Z_{i(2j)}>0.5\} \mathds{1} \{ Z_{i(2j+1)}>0.5\}\Big) A_i,
\end{align*}
where $p=25$, $\gamma_1=0.5$, $\gamma_2 = 0.99$ and $\gamma_3 = 0.33$. We use all combinations of the function types $f_R$ and $f_\tau$, which ensure that $f_\tau$ is less complex than $f_R$. 
This means, we only consider combinations, where $p_\tau \leq p_R$ and where $f_\tau$ is nonlinear only if $f_R$ is nonlinear as well. This restriction accounts for the fact that the CATE is typically less complex than the nuisance risk functions $\mu_{(1)}$ and $\mu_{(0)}$.

Based on the DGPs presented in Sections \ref{sec:4.1.2}--\ref{sec:4.1.4}, we examine the effect of varying $\theta$, $\gamma_1$, $e(\cdot)$ and the censoring rate under different complexities of the functions $f_R$ and $f_\tau$, while fixing $p_R = 25$ and $p_\tau=1$.

Regardless of the values assigned to $p_R$ and $p_\tau$, all 25 covariates are always used in model fitting. This implies that even in very sparse settings, where only one covariate has an effect on the hazard, learners have to handle a large number of nuisance variables. 

If $f_R(\boldsymbol{Z})$ or $f_\tau(\boldsymbol{Z}, A)$ were linear combinations of $\boldsymbol{Z}$ and $A$, the all-cause hazard would coincide with the hazard of a Cox-Weibull model. In that case, meta-learners employing Cox regression to estimate cumulative incidence functions are expected to outperform those relying on RSFs due to correct risk specification. In our simulation study, whenever the functions $f_R$ and $f_\tau$ are both categorized as linear, only the interaction terms between treatment and covariates prevent the correct specification of the hazard in the S-Learner, since we do not include any interactions in the Cox models. In contrast, the T-, X-, R- and DR-Learner model the absolute risk separately for each treatment arm, leading to correctly specified risk models when Cox (Lasso) regression is used.

\subsubsection{Varying Strength of Treatment Heterogeneity} \label{sec:4.1.2}
\citet{xu:2023:handbook} argue that the estimation of treatment effects becomes easier as the treatment effect heterogeneity increases. 
This is further implied by their simulation results which indicate that predictive performance declines for every estimator when the theoretical treatment heterogeneity is small. 
Moreover, methods misaligned with the functional form of the risk achieved a comparable performance to those accurately specifying the risk under larger treatment heterogeneity.  

We vary the strength of treatment heterogeneity through the parameter $\gamma_1$ in the function $f_\tau$, as defined in \Cref{sec:4.1.1}. A larger positive value of $\gamma_1$ leads to greater differences in the all-cause hazards between the treatment groups and thus increases treatment heterogeneity when measured on the log-hazard scale. Following \citet{xu:2023:handbook}, we also measure the treatment heterogeneity on the absolute risk scale through the ratio $\text{sd}_\tau / \text{sd}_{\mu_{(0)}}$ of the empirical standard deviation of the true CATEs $\text{sd}_\tau $ and the empirical standard deviation of baseline risks $\text{sd}_{\mu_{(0)}}$. Standardizing by 
$\text{sd}_{\mu_{(0)}}$ relates variation in treatment effects to inherent variation already present in baseline risks. Both standard deviation are computed from empirically generated true CATEs and baseline risks. 

\Cref{Tab:TH} displays the mean ratio over 100 Monte Carlo repetitions with sample size 5,000 that emerge from the systematic variation of $\gamma_1$. 
For almost every combination of $f_R$ and $f_\tau$, a larger value of $\gamma_1$ also yields a substantially higher treatment heterogeneity when measured using the absolute risk difference specific measure in \Cref{Tab:TH}. Only in the case, where $f_R$ and $f_\tau$ are both nonlinear, the empirical treatment heterogeneity on the absolute risk scale slightly decreases when $\gamma_1$ is increased from 0 to 0.5. Since the heterogeneity is nonetheless monotonically increasing with $\gamma_1$ in terms of the log hazard difference, this observation does not alter the qualitative comparison of meta-learners across heterogeneity settings. 

\begin{table}[ht] 
\centering
\caption{Treatment heterogeneity measured as the ratio between sd($\tau$), the empirical standard deviation of the true CATE, and sd($\mu_{(0)}$), the empirical standard deviation of the baseline risk. 
} 
\label{Tab:TH}
\begin{tabular}{c|c|c|c|c|c}  $f_R$ & $f_\tau$ & $p_R$ & $p_\tau$ & \multicolumn{1}{c|}{$\gamma_1$} & $\text{sd}_\tau /\text{sd}_{\mu_{(0)}}$ \\   \hline
    linear & linear & 25 & 1& 0.0 & 0.201 \\ 
 linear & linear & 25 & 1&  0.5 & 0.549 \\ 
 linear & linear  &25 & 1&  1.0 & 0.967 \\ 
nonlinear & linear  &25 & 1&  0.0 & 0.481 \\ 
 nonlinear & linear  & 25 & 1& 0.5 & 1.259 \\ 
 nonlinear & linear & 25 & 1& 1.0 & 2.056 \\ 
  nonlinear & nonlinear & 25 & 1& 0.0& 0.481 \\ 
   nonlinear & nonlinear & 25 & 1&  0.5 & 0.440 \\
    nonlinear & nonlinear & 25 & 1& 1.0 & 0.904 \\ 
   \hline
\end{tabular}
\end{table}

\subsubsection{Varying Treatment Assignment Procedure}  \label{sec:4.1.3}
We assess the predictive performance of the meta-learning algorithms under different treatment assignment mechanisms. In the previous settings, the treatment was assigned at random by a Bernoulli experiment with $A_i \thicksim \text{Bernoulli}(e(\boldsymbol{Z}_i))$, where $e(\boldsymbol{Z}_i) = 0.5$. In addition to this balanced assignment, we also consider unbalanced treatment assignment with a constant propensity score $e(\boldsymbol{Z}_i)=0.08$. Moreover, we explore settings with covariate-dependent treatment assignment, where the standard logistic regression model is misspecified. Under covariate-dependent treatment, we define the propensity score for an individual with covariates $\boldsymbol{Z}_i$ as 
\begin{align*}
e(\boldsymbol{Z}_i) = \dfrac{\exp(\delta(\boldsymbol{Z}_i))}{1+\exp(\delta(\boldsymbol{Z}_i))}, \quad\delta(\boldsymbol{Z}_i) = 2Z_{i1} Z_{i2} +2(\mathds{1}\{Z_{i3}>0\} - 1) + 3\mathds{1}\{|Z_{i4}| > q_{0.75}\} + 1.5(Z_{i4}^2 - 1),
\end{align*} 
where $q_{0.75}$ is the 0.75 quantile of the standard normal distribution. This setting is balanced in expectation in the sense that 
$E[e(\boldsymbol{Z}_i)]=0.5$, leading to treatment groups of similar size. 

In practice, covariate-dependent treatment assignment might occur when treatment is prioritized based on patient characteristics. In contrast, unbalanced treatment assignment may be the result of scarcity of a medication.

When $e(\boldsymbol{Z})\equiv0.50$ or $e(\boldsymbol{Z}) \equiv 0.08$, the logistic regression models for estimating the propensity score are correctly specified, which should benefit the X-, M-, R- and DR-Learner. The DR-Learner's doubly-robust property is expected to be advantageous even when the standard logistic regression model is misspecified (cov.-dep. setting). However, under unbalanced treatment assignment, the prediction performance of the DR-Learner may suffer due to the cross-fitting approach leading to smaller training datasets. This applies in particular to the models estimating $\mu_{(0)}$, which solely rely on observations from treatment arm 0.

\subsubsection{Varying Event and Censoring Rate}  \label{sec:4.1.4}
In practice, the event type distribution and the censoring rate are not under the investigator's control. High degrees of censoring result in a smaller effective sample size, decreasing the amount of information available to survival models. Similarly, if the event of interest is rare, a CATE estimation method which can handle few observations of the event of interest should be used.

We assess the effect of the proportion and therefore the number of type 1 events on CATE estimation by varying the parameter $\theta$. It follows from \Cref{proportion} that $\theta$ corresponds to the average expected proportion of type 1 events, without censoring being applied. We set $\theta$ to either $1/2, 1/3$ or $2/3$. 

To investigate the meta-learners under different degrees of right-censoring, we vary the scale parameter $\lambda$ of the Weibull distribution governing the censoring times and set $\lambda\in\{0.1, 0.3, 1.0\}$. In the settings with linear $f_R$, this corresponds to censoring rates of approx. 20\%, 35\% and 50\%, respectively. In both cases with nonlinear $f_R$, the mean censoring rate is about 10--15\% smaller in comparison across all values of $\lambda$. The exact realized censoring levels are displayed in \Cref{tab:cens} in Appendix \ref{sec:A.configs}. 

\subsection{Evaluation Criteria} \label{sec:4.2}
In line with \citet{xu:2023:handbook}, we evaluate and compare the meta-learners with regard to their ability to accurately estimate the CATE using the rescaled root mean squared error (RRMSE) and Kendall's $\tau_b$ \citep{kendall:1945:ties} as error metrics. 

The RRMSE is given by 
\begin{align*}
\text{RRMSE} = \dfrac{\sqrt{(\expval{(\widehat{\tau}(\boldsymbol{Z})-\tau(\boldsymbol{Z}))^2}}}{\text{SD}(\tau(\boldsymbol{Z}))},
\end{align*}
where $\widehat{\tau}(\boldsymbol{Z})$ and $\tau(\boldsymbol{Z})$ represent the predicted and the true CATE for covariates $\boldsymbol{Z}$, respectively. Without dividing by the standard deviation of the true CATE denoted by $\text{SD}(\tau(\boldsymbol{Z}))$, the quantity would correspond to the more frequently used root mean squared error. However, the division ensures that the performances of the meta-learners can also be compared across different settings of the simulation study, where the distributions of the true CATEs differ. For a finite test sample of $n$ observed covariate vectors $\boldsymbol{z}_1,\ldots,\boldsymbol{z}_n$, we calculate the RRMSE empirically as 
\begin{align*}
 \sqrt{\dfrac{ \frac{1}{n} \sum\limits_{i=1}^n (\tau(\boldsymbol{z}_i)- \widehat{\tau}(\boldsymbol{z}_i))^2}{\frac{1}{n-1} \sum\limits_{i=1}^n (\tau(\boldsymbol{z}_i) - \widebar{\tau} )^2  } },
\end{align*}
where $\widebar{\tau} = \frac{1}{n} \sum\limits_{i=1}^n \tau(\boldsymbol{z}_i)$. The RRMSE can be interpreted as the ratio between the variation not explained by the model and the variation present in the true conditional average treatment effects. 
Thus, an RRMSE value close to 0 is desirable. Moreover, neglecting the factor $\frac{n-1}{n}$ inside the square root, an RRMSE value greater than 1 suggests that the trained model provides worse predictions than the mean CATE model based on the actual CATEs (which is not available in practice).

Kendall's $\tau_b$ is a rank-based measure of association that allows for the assessment of the CATE model's treatment prioritization quality. This is especially relevant when a treatment is scarce or in possibly life-threatening situations. The $\tau_b$ statistic extends the original rank coefficient by \citet{kendall:1938:new} to settings with ties in either ranking. In the present context, it compares the ranking within the true CATE values $\tau(\boldsymbol{z}_1),\ldots,\tau(\boldsymbol{z}_n)$ with the ranking within the predicted CATE values $\widehat{\tau}(\boldsymbol{z}_1),\ldots,\widehat{\tau}(\boldsymbol{z}_n)$. For each pair of observations $(i,j)$, define a score based on the ground-truth CATE values as 
\begin{equation*}
    a_{ij} = \begin{cases}
        \phantom{-}1, \qquad \text{if $\tau(\boldsymbol{z}_i) < \tau(\boldsymbol{z}_j)$,}\\
        \phantom{-}0, \qquad \text{if $\tau(\boldsymbol{z}_i) = \tau(\boldsymbol{z}_j)$,}\\
        -1, \qquad \text{if $\tau(\boldsymbol{z}_i) > \tau(\boldsymbol{z}_j)$,}
    \end{cases}
\end{equation*} and define $b_{ij}$ analogously by replacing $\tau(\cdot)$ with $\widehat{\tau}(\cdot)$. Then, Kendall's $\tau_b$ is defined as \begin{align*}
    \tau_b = \dfrac{\sum\limits_{i=1}^n \sum\limits_{j=1}^n a_{ij} b_{ij}}{\sqrt{\left(\sum\limits_{i=1}^n \sum\limits_{j=1}^n a_{ij}^2\right) \cdot \left(\sum\limits_{i=1}^n \sum\limits_{j=1}^n b_{ij}^2\right)}} \in [-1,1],
\end{align*} where $\tau_b=1$ indicates that all pairs are ranked correctly, corresponding to perfect treatment allocation, whereas $\tau_b=-1$ indicates a fully reversed ranking. When predictions are constant, $\tau_b$ is undefined due to a division by zero. In that case, we argue to set $\tau_b=0$, since a constant CATE neither provides helpful nor misleading information in terms of treatment allocation. We provide occurrence rates thereof for all simulation settings and meta-learners in Appendix \ref{sec:A.further}.

We compute all evaluation metrics across 100 Monte Carlo replications for each simulation settings. In the following section, we visualize the distribution of these metrics using boxplots, displaying the 0.1, 0.25, 0.5, 0.75 and 0.9 quantiles. 

\section{Simulation Results} \label{sec:5}

Given the large number of learning algorithms considered (26 in total), we restrict the main presentation to a subset of results. In particular, we do not visualize the performances of the M-Learners and of the X-, R- and DR-Learners when combined with Cox regression without Lasso penalty. The M-Learner is excluded from the main analysis since it is consistently outperformed by the R-Learner. This finding is consistent with the results of \citet{xu:2023:handbook}. 

Instead, we refer to Appendix \ref{sec:A.further} where additional figures visualizing all learners at once are provided for both evaluation metrics. Given the presence of 25 covariates, applying Lasso regularization in Cox models is generally advisable. Therefore, we focus on the Cox Lasso for the X-, R- and DR-Learners within this section. The influence of the different data-generating settings on the risk models is still visible through the performance of the respective S- and T-Learners. Moreover, we focus on the RRMSE in the main analysis as the main evaluation metric. Boxplots for Kendall's $\tau_b$ are provided in Appendix \ref{sec:A.further} and are referred to in the discussion where relevant.

\subsection{Results: Varying Complexity of the Hazard Functions} \label{sec:5.C}
\begin{figure}[h]
    \centering
    \includegraphics[width=0.99\linewidth]{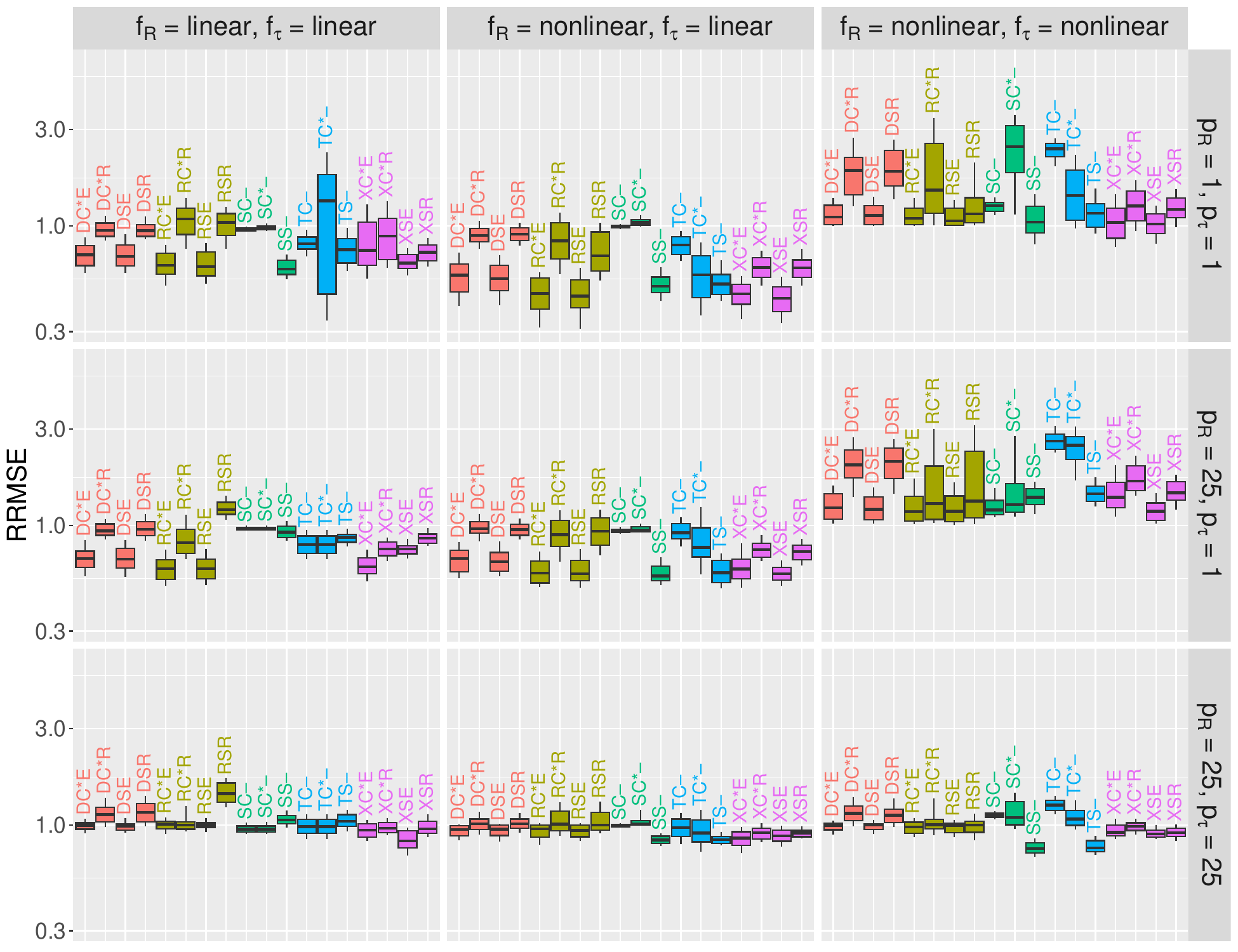}
    \caption{Boxplots showing the RRMSE of meta-learners for varying complexity of the hazard functions. 
    The parameters $\theta$ and $\gamma_1$ are both consistently set to 0.5 while $e(\boldsymbol{Z}) \equiv 0.5$ and $\lambda=0.1$.}
    \label{fig:5.1}
\end{figure}

\Cref{fig:5.1} displays the RRMSEs of the meta-learning algorithms under varying complexity of the hazards through boxplots. For clarity, we divide the analysis into multiple paragraphs. 

\textbf{Linear-Linear Setting}~~~The T-, X-, R- and DR-Learners model the absolute risk separately for each treatment arm. In the linear-linear case, their risk models are in consequence correctly specified when Cox (Lasso) regression is used. Thus, the TC*-- and TC-- are correctly specified in this setting. Unlike in the findings of \citet{xu:2023:handbook} on regular survival data, the SC*-- and TC*-- do not dominate this setting in our study. Instead, the R-Learners with elastic net CATE models achieve the best overall performance in the simple linear-linear setting, only when all 25 covariates influence the CATE function, the elastic net X-Learners become superior. 

In the special case of $p_R=p_\tau=25$, the TC*-- and TC-- exhibit the best performances with respect to $\tau_b$. However, we note that TC*-- is highly unstable in the sparse linear-linear setting ($p_R=p_\tau=1$) indicated by the large range in terms of RRMSE. This is likely due to regularization bias, as its non-penalized counterpart, TC--, is much more stable. Among the T-Learners, TS-- performs best in this case appearing far less sensitive to regularization bias than TC*--.

\textbf{Nonlinear-Linear Setting}~~~In scenarios with few covariate effects ($p_R=p_\tau=1$), the DR-, R- and X-Learners employing the elastic net and the SS--, TC*-- and TS--  perform better when $f_R$ is nonlinear rather than linear (given linear $f_\tau$). This also applies to the SC*--, SS-- and TS-- when more covariates influence the baseline risk ($p_R=25$, $p_\tau=1$). A potential reason for this behavior is that the higher treatment heterogeneity present in the nonlinear-linear case facilitates CATE estimation (see \Cref{Tab:TH} and \Cref{sec:5.TH}).

\textbf{Nonlinear-Nonlinear Setting}~~~Overall, predictive performance deteriorates when $f_\tau$ becomes nonlinear, reflecting the increased complexity of the CATE function. In the high-dimensional CATE setting ($p_\tau=25$) however, this effect is much less pronounced, suggesting that the prediction task is already similarly challenging even under linear $f_\tau$. Interestingly, the SS-- and TS-- even produce better predictions under nonlinear $f_\tau$ and $p_\tau=25$, although we don't have a theoretical explanation for this. In contrast, only increasing the complexity of the baseline risk function $f_R$ to nonlinear has comparatively little effect on most learners. 

In the nonlinear-nonlinear classical survival settings considered by \citet{xu:2023:handbook}, forest-based CATE models outperformed the considered Lasso regression approaches, whereas in linear-linear settings, the Lasso yielded the best performance. In our competing risks setup, however, the elastic net approaches (combining Lasso and Ridge regression) generally outperform their forest-based counterparts across all complexities. A plausible explanation is the reduction in effective sample size due to competing events as well as methodological differences (e.g., tuning and competing-risks-specific methods) compared to the original study.

\textbf{Effect of Dimensionality}~~~Increasing the amount of covariates that affect $f_R$ has little influence on performance. However, when the amount of covariates affecting the CATE through $f_\tau$ are increased, the relative performances of the learning algorithms change substantially. In these high-dimensional settings, we find that the X-Learners with elastic net perform most well when the hazard function is simple, whereas with increasing general hazard complexity, SS-- and TS-- become better alternatives. We further note that the performance of SC*-- and SC-- is close to the oracle mean model, but they become less stable across all complexities when $f_\tau$ becomes nonlinear.

\textbf{Comparison of Risk and CATE Models}~~~Our results also confirm that R-Learners (and DR-Learners) with elastic net are robust to the choice of absolute risk model, coinciding with the findings of \citet{xu:2023:handbook} for standard survival data. In contrast, the S- and T-Learner are sensitive to the choice of risk model. In our study, the S- and T-Learner consistently provide more stable performance when employing survival forests opposed to Cox (Lasso) regression. The Cox (Lasso) is only superior when the inherent proportional hazards assumption is not misspecified. 

\textbf{Recommendation}~~~When many covariates influence the CATE, X-Learners with elastic net or the RSF-based S- and T-Learner yield the best performances. Outside the high-dimensional CATE settings, R- and X-Learners with elastic net are reliable default choices. 

\textbf{Additional Results}~~~Figures \ref{fig:51aRRMSE} and \ref{fig:51aTAU} in Appendix \ref{sec:A.further} present boxplots summarizing the performance of all learners, including the previously excluded learners, for the evaluation metrics RRMSE and Kendall’s $\tau_b$, respectively. Overall, the conclusions drawn based on RRMSE remain qualitatively unchanged when considering $\tau_b$. \Cref{fig:51aRRMSE} further highlights that M--R generally yields inaccurate predictions, while the overall performance of M--E is slightly inferior to that of its R-Learner counterparts. Moreover, we observe that R-Learners as well as elastic net-based DR- and M-Learners frequently yield a constant CATE model in settings where both $f_R$ and $f_\tau$ are nonlinear (see \Cref{tab:acctau} in Appendix \ref{sec:A.further}), further indicating potential difficulties in capturing complex heterogeneity.

\subsection{Results: Varying Strength of Treatment Heterogeneity} \label{sec:5.TH}
\begin{figure}[hb]
    \centering
    \includegraphics[width=0.99\linewidth]{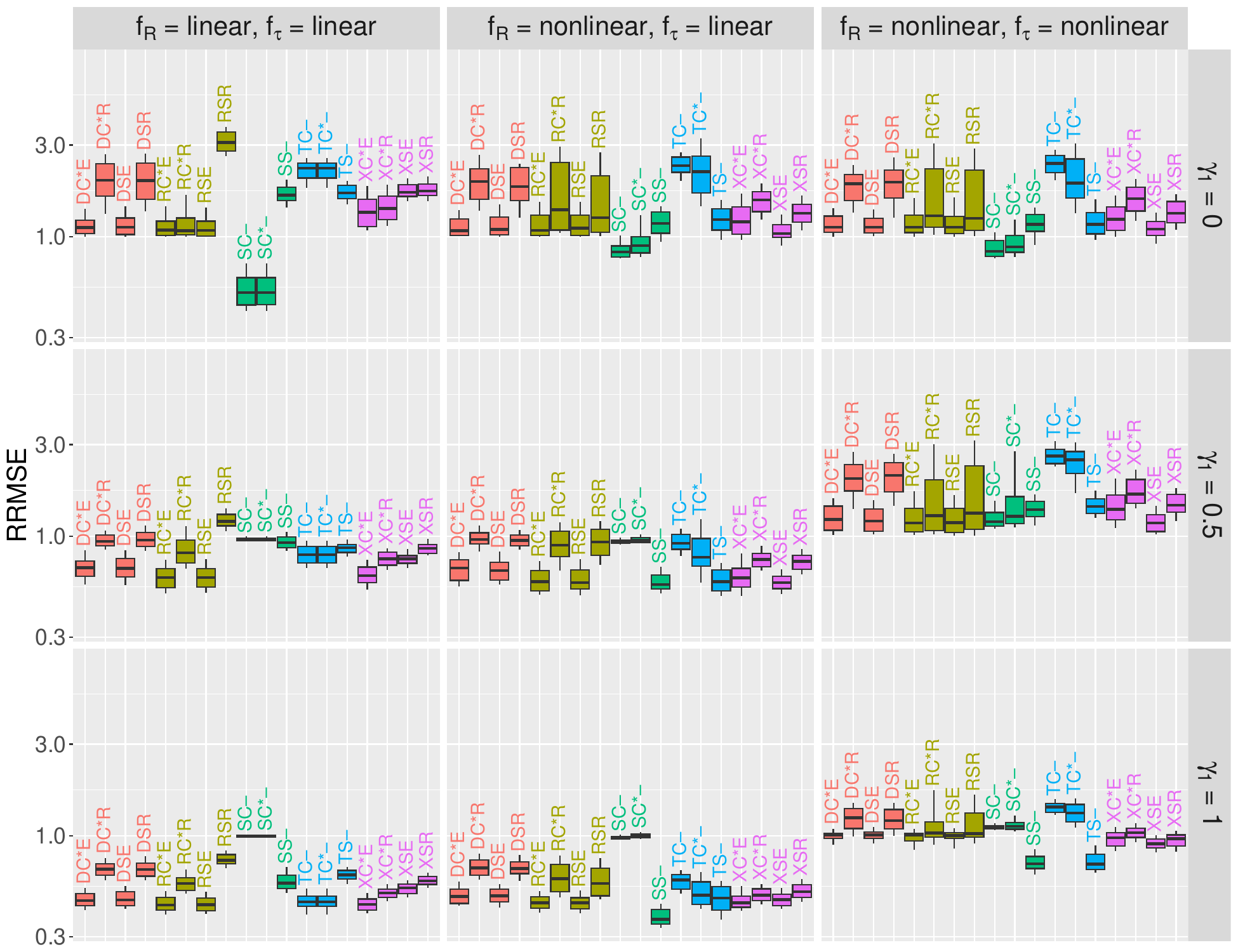}
    \caption{Boxplots showing the RRMSE of meta-learners for varying strength of treatment heterogeneity. A higher positive value of $\gamma_1$ indicates higher treatment heterogeneity. 
    The remaining parameters are set according to $p_R=p_\tau=1$, $\theta= 0.5$, $e(\boldsymbol{Z}) \equiv 0.5$ and $\lambda=0.1$. }
    \label{fig:5.3}
\end{figure}

\Cref{fig:5.3} depicts the boxplots of the RRMSE values for the considered selection of meta-learning algorithms across different strengths of treatment heterogeneity. The strength is regulated through the parameter $\gamma_1$, where a large $\gamma_1$ implies larger heterogeneity. The performances of all learners with respect to RRMSE and $\tau_b$ are depicted in Figures \ref{fig:53aRRMSE} and \ref{fig:53aTAU} in Appendix \ref{sec:A.further}, respectively. 

\textbf{Effect of Increasing Treatment Heterogeneity}~~~It is readily seen from \Cref{fig:5.3} that the RRMSEs decrease across almost all meta-learners with increasing $\gamma_1$. This decrease in the RRMSE values is accompanied by a reduction in their variance, as indicated by the substantially smaller boxplots. Hence, as anticipated, CATE estimation becomes more accurate with increasing $\gamma_1$ for most learners, which is in line with the results for the traditional survival setting \citep{xu:2023:handbook}. Overall, the relative rankings of the learners with respect to both RRMSE and Kendall’s $\tau_b$ remain largely unchanged as $\gamma_1$ increases.

\textbf{Performance Under Small Treatment Heterogeneity}~~~
The main exception to this trend are the SC*-- and SC--, which perform best under small treatment heterogeneity. Across most settings, they achieve RRMSE values close to the hypothetical oracle mean model. In the linear-linear setting with $\gamma_1=0$, they even substantially outperform all competing learners as well as the oracle mean model. This is intuitive, since in that setting the underlying hazard structure perfectly matches a Cox-Weibull model, rendering the SC*-- and SC-- correctly specified. In contrast, the TC*-- and TC-- are required to learn almost the exact same function under both treatment arms, which appears less efficient, although they are equivalently well-specified in this linear-linear setting. Apart from these learners, the elastic net-based R- and DR-Learners constitute reliable alternatives under small heterogeneity, although their performances with respect to $\tau_b$ are substantially worse to those of SC*-- and SC-- under $\gamma_1=0$.

\textbf{Comparison of CATE Modeling Approaches}~~~When increasing $\gamma_1$ the formerly best models SC*-- and SC-- get worse and are no longer the best approaches for any of the three settings. In fact, they perform even the worst for $\gamma_1=1$ when a linear $f_\tau$  is present (first two columns).  In comparison, forest-based CATE models become more competitive relative to their elastic net counterparts, as the performance gap decreases under stronger heterogeneity. This suggests that the effectiveness of the two CATE modeling strategies depends on the strength of treatment heterogeneity. Nevertheless, elastic net remains the generally preferred approach.

\subsection{Results: Varying Treatment Assignment Procedure}
\begin{figure}[h]
   \centering
   \includegraphics[width=0.99\linewidth]{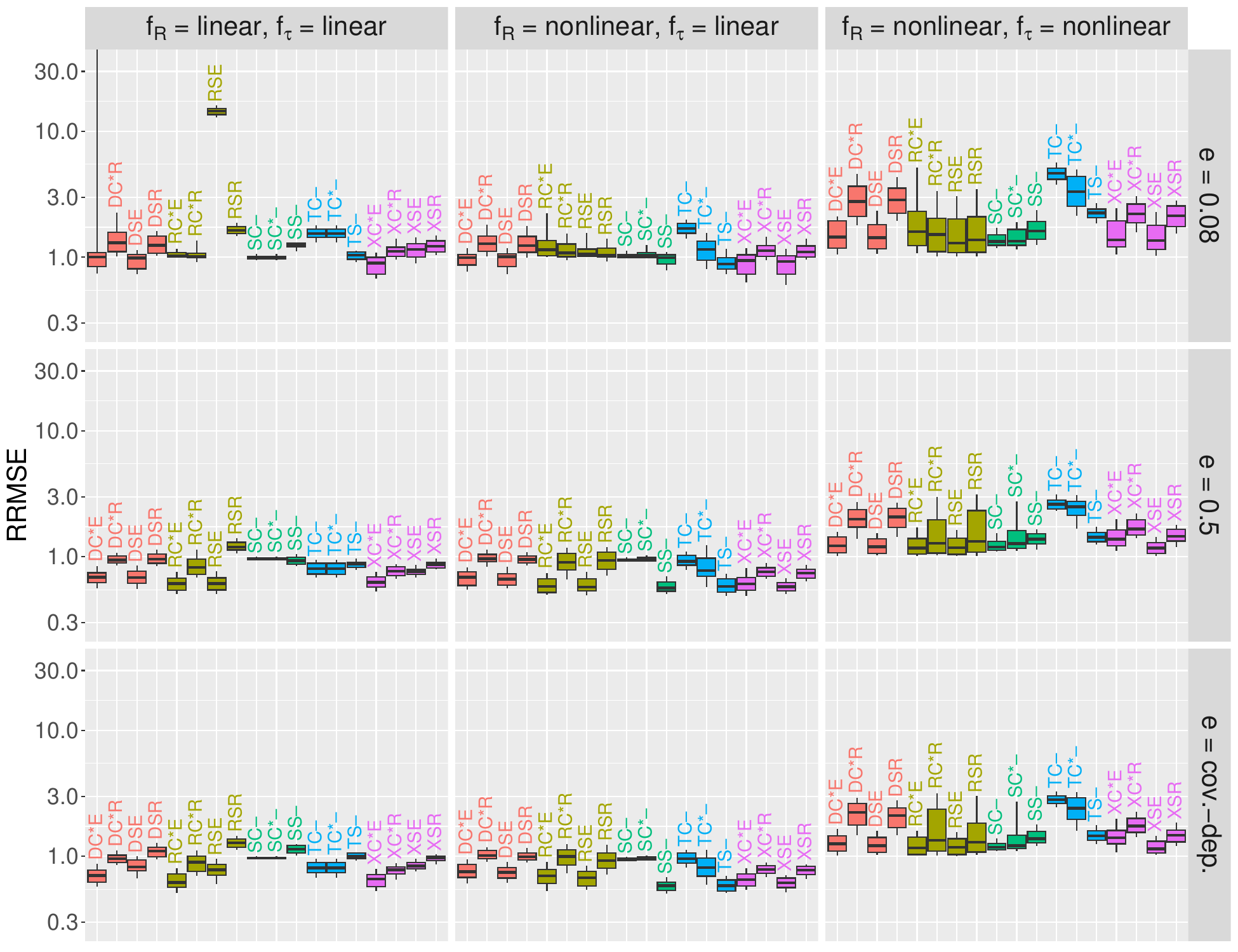}
   \caption{Boxplots showing the RRMSE of meta-learners for varying treatment assignment, expressed by the propensity score $e$.
    The remaining parameters are set according to $p_R=p_\tau=1$ and $\theta= \gamma_1=0.5$ and $\lambda=0.1$.}
    \label{fig:5.4}
\end{figure}

In this section, we evaluate the performance of the meta-learners under both balanced and unbalanced treatment assignment. Additionally, we investigate a setting with balanced but covariate-dependent treatment, in which the standard logistic regression model assumed within the applied X-, M-, R- and DR-Learners is misspecified. \Cref{fig:5.4} depicts the boxplots of the RRMSE values for the considered subset of meta-learners across these scenarios. The performances of all learners with respect to RRMSE and $\tau_b$ are presented in Figures \ref{fig:54aRRMSE} and \ref{fig:54aTAU} in Appendix \ref{sec:A.further}, respectively.

In the linear-linear setting with unbalanced treatment ($e=0.08$), extremely large RRMSE values are observed for several DR-Learners. Specifically, DC*E, DCE, DC*R, and DCR 
exhibit very large RRMSE values (up to $3.52 \cdot10^{60}$) in some instances
To ensure comparability of the remaining methods, the 0.9-quantiles of these learners are therefore not within the depicted axis range. These extreme values originate from instability in the Cox models, which is caused by the small number of treated units which becomes even smaller within the cross-fitting step. As a result, some predicted risks fall far outside the interval $[0,1]$, leading to CATE estimates well beyond the admissible range $[-1,1]$. Apart from this setting, the DC*E and DCE exhibit competitive performance. However, in practice, care must be taken to ensure stability of the underlying Cox models or more robust approaches such as the DSE or other learners may be preferred.

\textbf{Effect of Treatment Frequency}~~~Under strong treatment imbalance ($e=0.08$), the performance of most learners in terms of RRMSE deteriorates compared to the balanced setting ($e=0.5$). The S-Learners using Cox (Lasso) constitute an exception, as their performance remains largely unaffected. The X-Learners exhibit only minor performance degradation and therefore appear relatively robust to treatment imbalance, which is consistent with previous findings \citep{kuenzel:2019:stx, xu:2023:handbook}. In spite of that, X-Learners combined with random forests perform poorly relative to most other approaches. Among all learners, the XCE and XC*E achieve the best overall performance in the unbalanced setting. The DSE yields comparable results in terms of RRMSE, but is less reliable in terms of $\tau_b$. Moreover, the RSE performs very poorly under $e=0.08$ in the linear-linear setting, which stands in contrast to its otherwise competitive performance, typically comparable to that of the RC*E and RCE. Moreover, similar to the settings with nonlinear $f_R$ and $f_\tau$ analyzed in \Cref{sec:5.C}, R-Learners and elastic net-based DR- and M-Learners frequently produce constant CATE models under strong treatment imbalance (see Appendix \ref{sec:A.further}), indicating difficulties in the identification of heterogeneity.

\textbf{Effect of Propensity Misspecification}~~~In the setting with misspecified propensity scores, the performances of most learners barely change compared to the balanced, covariate-independent setting. This is expected for the S- and T-Learner, as they do not depend on a propensity score model. For the X-, M-, R- and DR-Learners with elastic net CATE models, a slight decrease in performance with regard to both metrics is observed under linear $f_\tau$. In contrast, their random forest-based counterparts barely change in performance with the notable exception of the M-Learner. Despite this robustness to misspecification of the propensity score, the random forest-based learners remain less accurate. The M--R is particularly sensitive to propensity score misspecification, as it exhibits substantially worse performance in both RRMSE and $\tau_b$. 

\subsection{Results: Varying Event and Censoring Rate}
The relative ranking of the considered meta-learners remains largely unchanged when varying either the event type distribution or the censoring rate with respect to both RRMSE and $\tau_b$. We therefore focus on the main observations and provide the corresponding boxplots for reference in Figures \ref{fig:52aRRMSE}, \ref{fig:52aTAU}, \ref{fig:55aRRMSE} and \ref{fig:55aTAU} in Appendix \ref{sec:A.further}.

As expected, increasing the censoring rate generally deteriorates predictive performance across learners, although the effect is less pronounced than the changes observed when varying the event type rate. Increasing the proportion of the event of interest, notably improves prediction accuracy across most learners. This is sensible, since a larger number of observed events of interest provides more information for model estimation. Another striking observation is that censoring itself has a stronger negative effect on all learners in the nonlinear-nonlinear compared to the two other settings. Moreover, while mostly being in line with the findings for the RRMSE, the performance difference between the different settings is more pronounced when measured in $\tau_b$.

Unlike \citet{xu:2023:handbook}, we did not compare different IPCW methods under varying censoring mechanisms and instead focused solely on the amount of censoring. Their results show that as one would expect, when censoring depends on covariates, Kaplan-Meier IPCWs suffer from misspecification and are thus outperformed by covariate-dependent random survival forest approaches. In contrast, in the covariate-independent setting the Kaplan-Meier approach outperforms the random survival forest due to its lower variance. We would expect the same logical results in the competing risks case, but further investigations are required.

\subsection{Summary of Results}
The simulation study highlights that the \textbf{complexity of the CATE function} has a substantial impact on predictive performance. Increasing treatment effect complexity through nonlinear interactions via $f_\tau$ leads to a strong deterioration in accuracy across most learners, whereas increasing the baseline hazard complexity via $f_R$ has only minor influence. In high-dimensional settings ($p_\tau=25$), the difference between linear and nonlinear $f_\tau$ becomes less prominent, suggesting that estimation is inherently difficult when many covariates contribute to treatment heterogeneity. Across all complexity settings, meta-learners employing elastic net for CATE estimation, particularly R- and X-Learners, demonstrate consistently strong and robust performance, while outperforming their random forest-based counterparts. When treatment effects depend on many covariates, the best performances are obtained by X-Learners with elastic net (under linear baseline risk) and RSF-based S- and T-Learners (under nonlinear baseline risk).

With respect to \textbf{treatment heterogeneity}, higher heterogeneity generally improves CATE estimation performance while preserving most relative rankings described in the previous paragraph. However, in the absence of heterogeneity, the Cox regression-based S-Learners perform best under this scenario, even better than under higher heterogeneity, due to correct model specification. 

Under \textbf{treatment imbalance}, performance deteriorates for most meta-learners. In particular, RSE and DR-Learners using Cox regression become unstable, whereas R-Learners based on Cox (Lasso) and elastic net regression (RCE and RC*E) remain comparatively stable, making them preferable within the class of R-Learners. Consistent with prior literature, X-Learners perform well under imbalance, with XCE and XC*E achieving the best results in these settings, although the random forest-based X-Learners generally underperform in comparison.

While lower censoring rates and higher rates of the event of interest generally lead to improved performance across most learners, the relative ranking of learners remains largely unchanged throughout the considered settings. Therefore, neither \textbf{event rates} nor \textbf{censoring} play a decisive role in the choice of an appropriate meta-learner.

Regarding \textbf{model components}, the choice of risk model has a limited impact for DR- and R-Learners and becomes important only under treatment imbalance. In contrast, X-Learners are more sensitive to the choice of risk model, and S- and T-Learners using random survival forests are more stable than their Cox-based counterparts. Moreover, across all learners that directly model the CATE, elastic net CATE models lead to better prediction performance than random forest models.

\textbf{Practical Recommendations}~~~Based on these findings, elastic net-based X-Learners and R-Learners using Cox regression can be recommended as robust default choices, due to their stability and overall performance across most settings. When the heterogeneity is expected to depend on many covariates, X-Learners with elastic net CATE modeling or survival forest-based S- and T-Learners (SS--, TS--), are competitive alternatives. In the presence of strong treatment imbalance, practitioners should prefer X-Learners with Cox (Lasso) and elastic net models (XCE and XC*E). 

\section{Illustrative Data Analysis - Hodgkin's Disease Study} \label{sec:6} 

In this section, we apply meta-learning using the \texttt{crsurvlearners} package to estimate CATEs on the Hodgkin's disease dataset available in the R package \texttt{randomForestSRC} (data \texttt{hd}, \cite{ishwaran:2023:rfsrc}). The Hodgkin's disease dataset contains records of $n=865$ Hodgkin's lymphoma patients that were treated at the Princess Margaret Hospital in Toronto between 1968 and 1986 \cite{competing:2006:pintilie}. All patients had either stage I or stage II Hodgkin's lymphoma. While 616 patients were treated with radiation alone (RT), the remaining 249 patients were treated with both radiation and chemotherapy (CMT). The data was recorded to analyze the long-term outcome in this group of patients with early stage Hodgekin’s lymphoma \cite{competing:2006:pintilie}. Hodgkin's lymphoma is a type of cancer that is ``\textit{most common in early adulthood (age 20–39 years) and in late adulthood (age 65 years and older)}'' \cite{hodgkin:2023:NCI}. When affected by the disease, malignant cancer cells develop in the lymph system.

We derive a CATE model that could be used in practice to assess which treatment, radiation alone or radiation combined with chemotherapy, is more beneficial for individual patients with early stage (stage I or stage II) Hodgkin's lymphoma. Specifically, we are interested in the long term effects of the treatments regarding the prevention of relapse, with death constituting the competing risk. While relapse in Hodgkin's lymphoma has been widely studied, only relatively few analyses have focused on very late relapses occurring 20 or more years after initial diagnosis \cite{liaskas:2026:focus}. Therefore, we model the CATE by the difference in the absolute risk 20 years after the first diagnosis. 

To select a suitable meta-learning algorithm for CATE estimation on the Hodgkin's disease dataset, the results of our simulation study provide a useful point of reference. However, since not all aspects of the simulated settings necessarily carry over to this real-data application, we additionally conduct a complementary simulation study based on data that more closely resemble the characteristics of the \texttt{hd} dataset. The generation of such synthetic competing risks data and the estimation of a CATE model can be conveniently implemented using the new \texttt{crsurvlearners} package, as we demonstrate in \Cref{sec:6.2}.

We describe the dataset in \Cref{sec:6.1} and discuss the model fitting procedure including the preceding mimicking study in \Cref{sec:6.2}. Based on the estimated CATE model, we identify population subgroups with heterogeneous group-internal average treatment effects in \Cref{sec:6.3} using a ``fit-the-fit'' approach.

\subsection{Data Description} \label{sec:6.1}

The dataset comprises the variables age, sex, treatment (\texttt{trtgiven}), mediastinum involvement (\texttt{medwidsi}), extranodal disease (\texttt{extranod}), clinical stage (\texttt{clinstg}), time and status. The variable names in parentheses correspond to the respective variable names in the dataset. If no variable name is provided in parentheses, the given variable name is the same as the one in the dataset. All 865 observations contain no missing values.

Each patient received either radiation alone (RT) or a combination of radiation and chemotherapy (CMT) as treatment. The time variable measures the time in years until the first occurrence of an event starting from the diagnosis. The status variable describes the event type, with 0 indicating censoring, 1 indicating the event of interest (relapse) and 2 indicating death without experiencing relapse. While 291 patients relapsed, 135 died without relapse. The remaining 439 observations are censored. 
The variable mediastinum involvement indicates the extent to which the mediastinum is affected by the cancer. It is either none, small or large. The variable extranodal disease indicates whether the disease is extranodal or nodal. \Cref{tab:variables}, which builds upon Table 3 from \citet{ruehl:2024:resampling}, provides an overview of the variables present in the dataset separated by treatment. For the categorical variables, the relative frequencies of the variable values in the respective treatment group are displayed in parentheses following the absolute frequencies. For the metric variables age and time, \Cref{tab:variables} instead shows the corresponding mean and standard deviation values.
\begin{table}[ht]
\renewcommand{\arraystretch}{1.15}
\caption{Overview of the Hodgkin's disease data per treatment group.}
\label{tab:variables}
\centering
\begin{tabular}{l|r|r} \hline
   Variable &  \multicolumn{2}{l}{\phantom{Ml Radiation alon}Treatment:} \\ 
 & \phantom{Ml} Radiation alone \phantom{Ml} & Radiation and chemotherapy  \\ 
 & \multicolumn{1}{c|}{($n=616$)} & \multicolumn{1}{c}{($n=249$)} \\
   \hline
   Sex: & & \\ 
   \phantom{M} male & 331 (53.73\%) \phantom{MM} & 132 (53.01\%) \phantom{herapy} \\ 
   \phantom{M} female & 285 (46.27\%) \phantom{MM} & 117 (46.99\%) \phantom{herapy} \\ 
   Clinical stage: && \\ 
   \phantom{M} stage I  & 266 (43.18\%) \phantom{MM} & 30 (12.05\%) \phantom{herapy} \\ 
   \phantom{M} stage II &  350 (56.82\%) \phantom{MM} & 219 (87.95\%) \phantom{herapy} \\ 
   Mediastinum involvement: & & \\ 
   \phantom{M} none & 382 (62.01\%) \phantom{MM} & 82 (32.93\%) \phantom{herapy} \\  
   \phantom{M} small &211 (34.25\%) \phantom{MM} & 77 (30.92\%) \phantom{herapy} \\  
   \phantom{M} large & 23 \phantom{4}(3.73\%) \phantom{MM} & 90 (36.14\%) \phantom{herapy} \\  
   Extranodal disease: & & \\  
   \phantom{M}nodal disease & 587 (95.29\%) \phantom{MM} & 199 (79.92\%) \phantom{herapy} \\  
   \phantom{M}extranodal disease: & 29 \phantom{4}(4.71\%) \phantom{MM} & 50 (20.08\%) \phantom{herapy} \\ 
   Status: && \\
   \phantom{M} censored & 293 (47.56\%) \phantom{MM} & 146 (58.63\%) \phantom{herapy} \\
   \phantom{M} relapse & 230 (37.34\%) \phantom{MM} & 61 (24.50\%) \phantom{herapy} \\
   \phantom{M} death & 93 (15.10\%) \phantom{MM} & 42 (16.87\%) \phantom{herapy} \\ \hline
   Age, mean (sd) & \multicolumn{1}{c|}{35.93 (16.37)} & \multicolumn{1}{c}{33.77 (12.86)} \\ 
   Time, mean (sd) & \multicolumn{1}{c|}{12.34 \phantom{1}(9.55)} & \multicolumn{1}{c}{14.16 \phantom{1}(8.54)} \\ \hline
\end{tabular}
\end{table}

To select a suitable meta-learning algorithm for CATE estimation on the Hodgkin's disease dataset, we may first draw on the findings of our simulation study. The results indicate that the amount of censoring and the event type distribution have only a minor impact on CATE estimation, as the relative performance of the considered learners remained largely unchanged across the corresponding settings. While the degree of treatment heterogeneity is generally unknown in practice, it may be informed by domain expertise. 

In accordance with the findings of \citet{ruehl:2024:resampling}, we can assume that the proportional hazards assumption, as defined in \Cref{eq:coxph}, is satisfied for both causes. 
We thus conclude that the linear-linear hazard setting is the most comparable. 

The treatment proportion in the Hodgkin's disease dataset is 28.79\%, which lies between the previously considered settings of 8\% and 50\%. Moreover, the observed censoring rate is close to 50\%. In consequence, the ideal learner should exhibit strong results in the linear-linear settings across all propensity score configurations and the simulation settings where $\lambda=1.0$ resulting in a comparable censoring rate. Taken together, these findings suggest the use of the X- or R-Learner with elastic net CATE model and Cox (Lasso) risk model.

We additionally conduct a mimicking study in which the simulated data more closely resemble the characteristics of the dataset at hand. In addition to the previously mentioned differences, the \texttt{hd} dataset contains fewer covariates, includes multiple categorical variables, and is substantially smaller than the training datasets used in our simulation study (865 vs. 5000 observations). In the following section, we demonstrate how such synthetic datasets can be generated using the \texttt{crsurvlearners} package. 

\subsection{Mimicking Study and CATE Estimation with the \texttt{crsurvlearners} Package} \label{sec:6.2}
\textbf{Data Generating Process}~~~To generate datasets with more similar characteristics to the \texttt{hd} dataset, we modify the DGP from the simulation study (\Cref{sec:4}) as follows. Since the meta-learners require numerical inputs, we consider the categorical variables in the form of binary variables resulting in 5 binary variables, since the mediastinum involvement requires 2 dummy variables. We generate the resulting binary variables from Bernoulli variables with the relative frequencies as success probabilities. The single continuous covariate age is represented by a random variable that follows a standard Gaussian distribution as in the base DGP of the simulation study. As the Gaussian permits negative values, we interpret 
this as a generic continuous predictor representing standardized age. In total, this yields $p=6$ covariates and we assume that each covariate may influence the baseline risk as well as the CATE, i.e., we set $p_R = p_\tau=6$. As previously argued, we restrict the data to the linear-linear setting.
The all-cause hazard is generated as in \Cref{eq:all-cause} which depends on the functions $f_R$ and $f_\tau$. We adapt these functions in the same style as in the simulation study to the setting of $p=p_R=p_\tau=6$ covariates. Given covariates $\boldsymbol{Z}_i$ and treatment $A_i$, these correspond to 
\begin{align*}
    f_R(\boldsymbol{Z}_i) &= \sum\limits_{j=1}^6 \beta_1 \dfrac{Z_{ij}}{\sqrt{p}}, \\
    f_\tau(\boldsymbol{Z}_i, A_i) & = \left(-0.5-\sum\limits_{j=1}^6 \gamma_1 \dfrac{Z_{ij}}{\sqrt{p}}\right)A_i,
\end{align*}
where $\beta_1=1$ and $\gamma_1=0.5$ control the treatment heterogeneity. In accordance with the \texttt{hd} dataset, we set the type-1 event rate to $\theta=0.68$, the treatment rate to $e=0.71$ and consider $n=865$ training observations. We further set $\lambda=1.5$ in the censoring Weibull distribution while keeping the shape at $\nu = 1$, thereby achieving a censoring rate of about 50\% (as in the \texttt{hd} dataset) given the other parameter settings. We further use the same truncation of survival times and $t_0=1.2$ as in the base DGP. 

While the cause-specific hazards and censoring distribution in this simulation setting are not intended to represent the unknown data-generating mechanisms of the \texttt{hd} dataset, the proposed design nevertheless matches several key characteristics of the observed data more closely. To select the most suitable meta-learner we create such training data and evaluate the RRMSE and Kendall's $\tau_b$ on 5000 test data points over 100 Monte Carlo repetitions. These customized datasets can be created using the \texttt{crsurvlearners} package via the below code snippet. We note that the specified data generating functions are restricted to 2 competing events, whereas the learner implementations allow for more competing events.

\begin{lstlisting}[style=Rstyle]
library(crsurvlearners)

## Specify data-generating mechanism
g <- specify_data(type_R = "linear", type_tau = "linear", p = 6, p_R = 6, p_tau = 6, 
                  p_bin = 5, pctgs_bin = c(0.54, 0.34, 0.13, 0.54, 0.09), 
                  theta = 0.68, gamma_1 = 0.5, e = 0.71,
                  c_scale = 1.5, c_shape = 1, t0 = 1.2, Y.max = 12,
                  n_true_CATE = 10000, n_train = 865, n_test = 5000
                  )
## Generate training and test data
D_train <- generate_data(g, train = TRUE)
D_test  <- generate_data(g, train = FALSE)
\end{lstlisting}

\begin{figure}[h]
    \centering
    \includegraphics[width=0.99\linewidth]{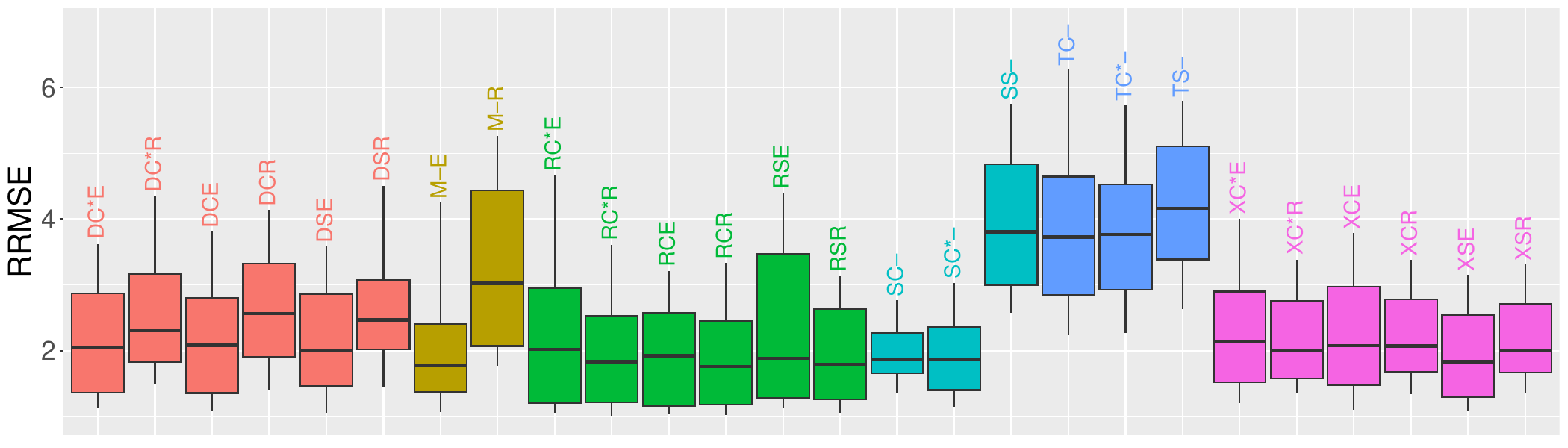}
    \caption{Boxplots showing the RRMSE of all meta-learning algorithms in the mimicking study.}
    \label{fig:mimicRRMSE}
\end{figure}

\begin{figure}[h]
    \centering
    \includegraphics[width=0.99\linewidth]{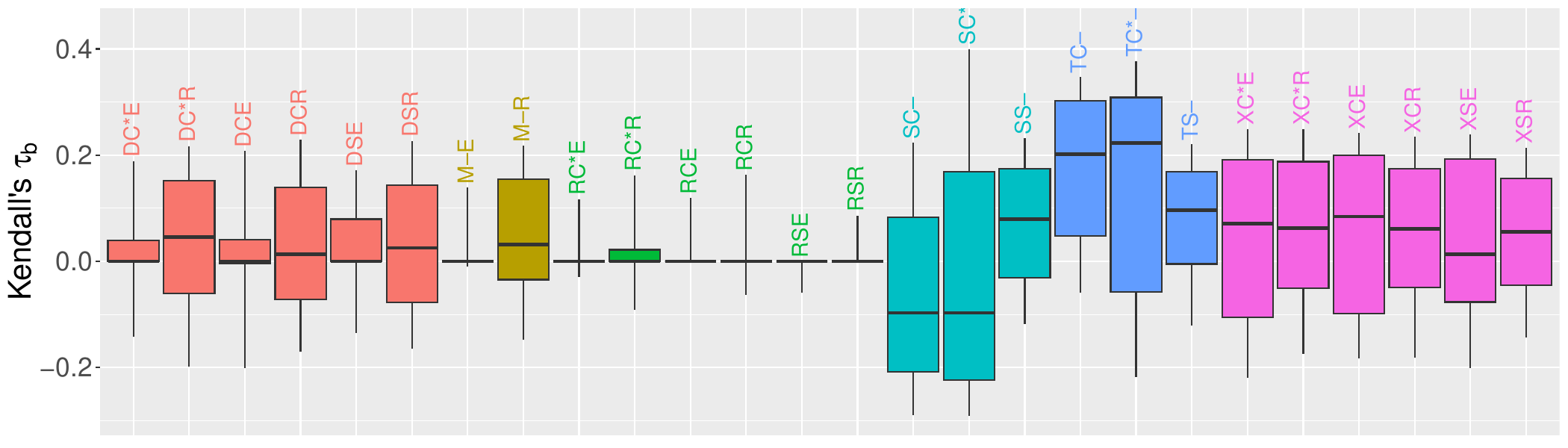}
    \caption{Boxplots showing Kendall's $\tau_b$ of all meta-learning algorithms in the mimicking study.}
    \label{fig:mimicTaub}
\end{figure}

\textbf{Simulation Results}~~~\Cref{fig:mimicRRMSE} and \Cref{fig:mimicTaub} show the performances of all considered meta-learners in terms of RRMSE and Kendall's $\tau_b$, respectively. Compared to the results from the simulation study, we generally find both larger RRMSEs and small absolute values of $\tau_b$ indicating that estimation in this small sample setting with substantial censoring is more challenging. 

RCR yields the smallest median RRMSE of 1.76 performing slightly better in that regard than M--E with a median RRMSE of 1.77. However, the M--E exhibits a much larger 0.9-quantile than RCR and thus should not be preferred. Overall, R-Learners with random forest CATE models yield good performances and outperform their elastic net counterparts, in contrast to the large-scale simulation study. This pattern is not reflected for the DR- and X-Learners. Apart from the R-Learners with random forests, the DR-Learners with elastic net CATE model, the SC--, SC*-- and most X-Learners show comparable results. In terms of $\tau_b$, the best performances in terms of the median are achieved by the TC-- and TC*-- with values slightly above 0.2. This highlights that most estimated models are rather weak and mostly do not provide reliable treatment allocation suggestions in this simulated setting. Interestingly, these two learners were among the worst performers in terms of RRMSE.

We decide to use the RCR to estimate our CATE model due to achieving the lowest median RRMSE. To use the RCR-Learner for CATE estimation, we assume non-informative censoring and make the necessary identifiability assumptions (see \Cref{sec:2.3}). The positivity assumption is addressed below, although a thorough investigation of it and all other remaining assumptions is not possible due to limited information about the data acquisition process. 

\textbf{CATE Model}~~~Using the \texttt{crsurvlearners} package, in the code snippet below the RCR model is estimated after transforming the dataset accordingly. With the \texttt{cr\_cate()} function any learner discussed in this work can be called by specifying the desired combination of meta-learner, risk model and CATE model. It also allows to pass further arguments for the executed supervised learning tasks including IPCW and propensity scores.

\begin{lstlisting}[style=Rstyle]
library(crsurvlearners)
library(randomForestSRC)
set.seed(18)

data("hd")
hd$trtgiven <- as.factor(ifelse(hd$trtgiven == "RT", 1, 0))
hd$sex <- ifelse(hd$sex == "M", 1, 0)
hd$clinstg <- ifelse(hd$clinstg == 1, 1, 0)
hd$extranod <- ifelse(hd$extranod == "Y", 1, 0)
hd$medwidsiL <- ifelse(hd$medwidsi == "L", 1, 0)
hd$medwidsiN <- ifelse(hd$medwidsi == "N", 1, 0)
Z <- hd[ ,c("age", "sex", "clinstg", "extranod", "medwidsiL", "medwidsiN")]

model <- cr_cate(Z, hd$trtgiven, time = hd$time, event = hd$status, t0 = 20, 
                 cause = 1, ml = "r", riskmod = "cox", catemod = "rf")
\end{lstlisting}
The treatment variable was encoded, such that 1 indicates radiation therapy and 0 indicates the combination of radiation and chemotherapy. As in the simulation study, IPCWs for the complete observations are computed in an out-of-sample manner based on 10 folds of the dataset. In treatment group 1, there are 456 complete observations, while 171 complete observations are present in treatment group 0. 

Within the RCR, a logistic regression with elastic net penalty is used to estimate the propensity score. We investigate the validity of this approach using a 10-fold cross-validation of a regular logistic regression on the full dataset. When classifying the observations based on a threshold of 0.5, the resulting mean accuracy equals 0.795. Furthermore, the mean area under the receiver operating curve (AUC) is given by 0.786 further justifying the use of a logistic regression. We further find that the resulting out-of-fold predictions for the propensity score all lie within the interval $[0.05,0.95]$ indicating that the positivity assumption is probably fulfilled.

The estimated CATE model predicts the difference between the two treatment groups in the absolute risk 20 years after the diagnosis conditioned on the observed covariates. We use the resulting in-sample predictions from the dataset in the following to identify how different subpopulations, specified by subregions of the covariate space, benefit from the treatments.

\subsection{Subgroup Analysis} \label{sec:6.3}

To identify subgroups with heterogeneous treatment effects, we fit a small regression tree using the CATE predictions from the RCR as the response and the considered covariates as predictors. While the RCR CATE model provides treatment effects within small subgroups specified by all covariates, this ``fit-the-fit'' approach summarizes treatment effects over broader subpopulations defined by the tree splits. The leaf nodes of the fitted tree represent average treatment effects within these subgroups. Such fit-the-fit approaches are commonly used to improve the interpretability of flexible models, such as Bayesian additive regression trees \citep{Logan:2019:ITEbarts, hu:2021:estimating, hu:2021:lungcancer}, and constitute an efficient way to derive meaningful interpretations about treatment heterogeneity \citep{hu:2021:lungcancer}. Similarly, in contrast to the less interpretable RCR model, which is composed of multiple different predictive models, the regression tree allows to explain different treatment effects across determined subgroups. 

\begin{figure}[h]
    \centering
    \includegraphics[width=0.99\linewidth]{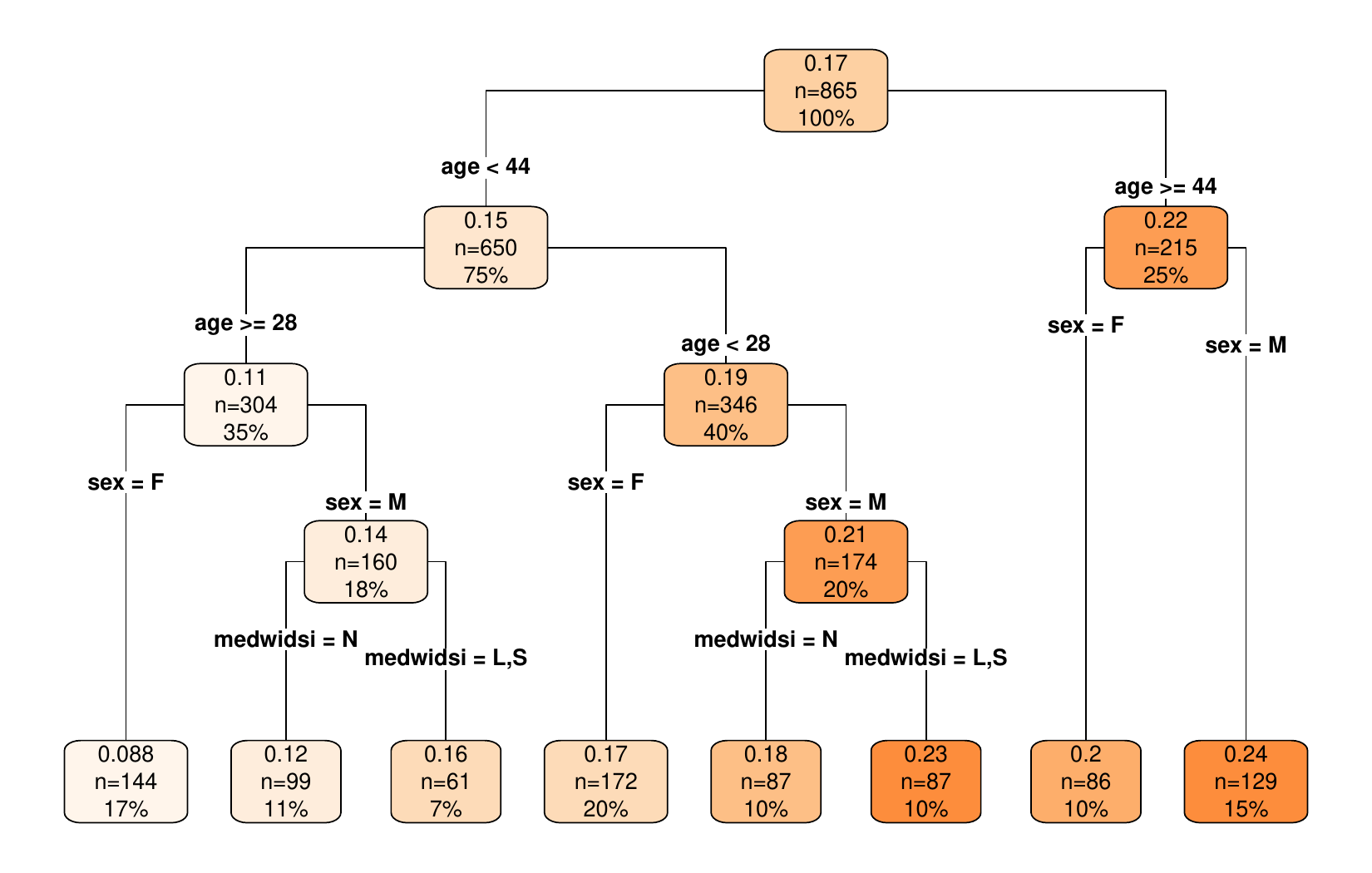}
    \caption{Regression tree relating covariates to CATEs predicted by the RCR. Each node shows the mean predicted CATE and the number of observations (in total and percentage) assigned to it.}
    \label{fig:RCRtree}
\end{figure}

\Cref{fig:RCRtree} illustrates the fitted regression tree. It identifies heterogeneous subgroups based on the variables age, sex and mediastinum involvement. Radiation therapy was encoded as treatment 1 and the combination of radiation and chemotherapy as treatment 0. Therefore, a CATE greater than 0 corresponds to a better prognosis in the prevention of relapse in a time frame of 20 years when treated with radiation and chemotherapy. In contrast, when the CATE is smaller than 0, radiation alone is more promising. In all determined subgroups, the group-internal average treatment effect is greater than 0, suggesting that the combination of radiation and chemotherapy is the better treatment across the whole population. In fact, all predicted CATEs of the dataset are truly greater than 0. This result is in line with current clinical practice, in which combined-modality treatment represents the standard care for early-stage Hodgkin lymphoma \citep{patel:2018:reduced}.

The averaged conditional treatment effect across the full population is estimated at 0.17. The first split of the tree separates patients aged below 44 years from those that are at least 44 years old. While the average CATE is 0.15 for patients younger than 44 years, it corresponds to 0.27 for patients aged 44 years and above. This indicates a substantially larger benefit of adding chemotherapy in older patients. 

Within the younger group, a further split at age 28 reveals that patients aged 28 to 43 years benefit the least from combined treatment with an average treatment effect of 0.11, compared to those aged below 28 with an average of 0.19. Across all age groups, subsequent splits by sex show consistently higher treatment effects for males, with differences ranging from 0.04 to 0.052, suggesting that male patients benefit more from combined therapy than females in the prevention of 20 year relapse.

Notably, these splits align with established prognostic factors from the International Prognostic Score (IPS) for advanced Hodgkin's disease \citep{hasenclever:1998:prognostic}, where age $\geq$ 45 at diagnosis and male sex are associated with poorer survival outcomes. This indicates that the RCR model identifies high-risk patients and suggests that the combination of chemotherapy and radiation therapy, the current standard care, provides especially strong benefits for these groups in terms of long-term relapse prevention.

Among both male groups younger than 44 years, an additional split is performed based on whether the mediastinum is affected by cancer or not. In both resulting subgroups with mediastinum involvement, patients exhibit a treatment effect increase by 0.04--0.05 than in the respective subgroup without involvement. As with the preceding splits on age and sex, the split rule is consistent with the role of mediastinum involvement as a risk factor for relapse \cite{velentjas:1980:mediastinal}.

Overall, the tree identifies eight subgroups with heterogeneous treatment effects, with age and sex constituting the main drivers. The largest benefit of additional chemotherapy is observed for males aged 44 years and above (avg. CATE 0.24), whereas females aged between 28 and 43 years benefit the least (avg. CATE 0.088). The findings suggest that the RCR model identifies well-known risk factors and estimates a larger benefit of the combined therapy to higher risk patients.

\section{Conclusion} \label{sec:7}
In this work, we studied the estimation of conditional average treatment effects (CATEs) from right-censored survival data with competing risks under binary treatment, using baseline covariates. Such CATE models can support treatment decision-making in personalized medicine, but their performance cannot be directly evaluated in practice. To provide guidance on model selection, we therefore considered a range of meta-learners and systematically evaluated their performance across diverse simulation settings, varying in hazard complexity, treatment heterogeneity, treatment assignment, censoring, and event type distributions. 

Our results show that the complexity of the underlying CATE function and the degree of treatment imbalance strongly influences the relative performance of meta-learners, whereas the amount of heterogeneity and event distribution have a minor effect. Based on our findings, elastic net-based R-Learners and X-Learners constitute strong default choices, with elastic net-based X-Learners being more robust to high treatment imbalance. Overall, elastic net-based CATE models outperform random forest-based alternatives in most of the simulated competing risks scenarios.

To support practical application, this work is accompanied by the R package \texttt{crsurvlearners}, which implements the discussed meta-learners and enables interpretable subgroup analyses and reproducible simulation studies. We demonstrated its applicability using real-world data on Hodgkin's lymphoma, comparing radiation therapy alone to a combined chemo-radiotherapy with regard to the prevention of relapse, and performed a subgroup analysis based on the estimated CATE model. The subgroup analysis suggests that the model identifies well-known clinical risk factors and assigns larger treatment benefits under the combined treatment to higher-risk patients.

In summary, this work provides methodological details on CATE estimation under competing risks and offers practical guidance for model selection. The presented findings rely on standard causal assumptions, which may limit their applicability where these are violated, for instance due to unmeasured confounding. Future work may investigate violations of these assumptions and extend this work by focusing on time-dynamic covariates, multiple states, multivariate treatment, and alternative CATE estimators beyond meta-learners.

\section*{Acknowledgements}
We gratefully acknowledge the computing time provided on the Linux HPC cluster at
Technical University Dortmund (LiDO3) which we used to conduct our simulation study.
LiDO3 is partially funded in the course of the Large-Scale Equipment Initiative by
the Deutsche Forschungsgemeinschaft (DFG, German Research Foundation) as project
271512359. Moreover, Markus Pauly was supported by the DFG project 352692197 and Sarah Friedrich was supported by the DFG project 439942859. We also like to thank Jasmin R\"uhl for helpful discussions. 
 The underlying Master's thesis \cite{klippert:2024} of the first author was awarded with the 2025 Bernd-Streitberg prize of the German Region of the International Biometric Society.


\bibliographystyle{unsrtnat-new}   
\bibliography{bib.bib}      

\newpage
\appendix

\section{Hyperparameter Optimization within Meta-Learners} \label{sec:A.hyperparams}
Within the considered meta-learners, hyperparameters are optimized for all supervised learning components, including risk, CATE, and propensity score models. We briefly outline the corresponding tuning procedures.

\textbf{Risk Models}~~~For Cox Lasso, the regularization parameter is selected via a 5-fold cross validation over a grid of candidate values within $[0.01, 100]$. In the case of the competing risks random survival forest, we use the built-in tuning procedure \texttt{tune.rfsrc} from the \texttt{randomForestSRC} R package \citep{ishwaran:2007:rsfR, ishwaran:2008:rsf, ishwaran:2023:rfsrc} to tune the mtry parameter and the minimum node size. This procedure performs a bisection search over the mtry parameter to minimize the out-of-sample error for fixed values of the minimum node size. The smallest out-of-sample error determines the optimal pair of parameters. To reduce computational cost, 100 trees are used during tuning, while the final models are fitted with 250 trees.

\textbf{Direct CATE Models}~~~For random forest-based CATE models, hyperparameters are tuned using the built-in procedure of the \texttt{regressionforest} function from the \texttt{grf} R package \cite{grf}. The sample fraction, minimum node size, and mtry parameter are optimized by minimizing the out-of-bag error over randomly sampled parameter configurations. The sample fraction parameter governs the amount of samples drawn for each tree. The intention behind adjusting the sample fraction is that more diverse trees might be achieved by using fewer samples in the creation of each tree. Diverse trees are less correlated, which is beneficial for reducing the variance of the random forest \cite{isla}. As before, the number of trees during tuning is set to 100, while the final random forest models consist of 250 trees. For elastic net regression, we employ a 5-fold cross-validation over 100 randomly sampled combinations of the penalty parameters.

\textbf{Propensity Models}~~~For propensity score estimation in the X-, M-, R-, and DR-learners, we use logistic regression with an elastic net penalty. The penalty parameters are tuned analogously via 5-fold cross-validation using the penalized log-loss as the objective function.

\newpage
\section{Simulation Configurations} \label{sec:A.configs}
\Cref{A:Tabelle} summarizes all simulation settings considered in the large-scale simulation study, along with the corresponding parameter specifications. In total, 33 distinct configurations were examined. 

\begin{table}[h] 
\centering \caption{Overview of the configurations of the data generating process used in the simulation study.} \label{A:Tabelle}
\begin{tabular}{c|c|c|r|r|r|r|c|c}
  \hline
Config. & $f_R$ & $f_\tau$ & \multicolumn{1}{c|}{$p_R$} & \multicolumn{1}{c|}{$p_\tau$} & \multicolumn{1}{c|}{$\theta$} & \multicolumn{1}{c|}{$\gamma_1$} & $e$ & $\lambda$ \\ 
  \hline
  1 & linear & linear & 1 & 1 & 1/2 & 0.5 & 0.50 & 0.1\\ 
    2 & linear & linear & 25 & 1 & 1/2 & 0.5 & 0.50 & 0.1 \\ 
    3 & linear & linear & 25 & 25 & 1/2 & 0.5 & 0.50 & 0.1 \\ 
    4 & nonlinear & linear & 1 & 1 & 1/2 & 0.5 & 0.50 & 0.1 \\ 
    5 & nonlinear & linear & 25 & 1 & 1/2 & 0.5 & 0.50 & 0.1 \\ 
    6 & nonlinear & linear & 25 & 25 & 1/2 & 0.5 & 0.50 & 0.1 \\ 
    7 & nonlinear & nonlinear & 1 & 1 & 1/2 & 0.5 & 0.50 & 0.1 \\ 
    8 & nonlinear & nonlinear & 25 & 1 & 1/2 & 0.5 & 0.50 & 0.1 \\ 
    9 & nonlinear & nonlinear & 25 & 25 & 1/2 & 0.5 & 0.50 & 0.1 \\  \addlinespace\midrule\addlinespace 
   10 & linear & linear & 25 & 1 & 1/3 & 0.5 & 0.50 & 0.1 \\ 
   11 & linear & linear & 25 & 1 & 2/3 & 0.5 & 0.50 & 0.1 \\ 
   12 & nonlinear & linear & 25 & 1 & 1/3 & 0.5 & 0.50 & 0.1 \\ 
   13 & nonlinear & linear & 25 & 1 & 2/3 & 0.5 & 0.50 & 0.1 \\ 
   14 & nonlinear & nonlinear & 25 & 1 & 1/3 & 0.5 & 0.50 & 0.1 \\ 
   15 & nonlinear & nonlinear & 25 & 1 & 2/3 & 0.5 & 0.50 & 0.1 \\  \addlinespace\midrule\addlinespace 
   16 & linear & linear & 25 & 1 & 1/2 & 0.0 & 0.50 & 0.1 \\ 
   17 & linear & linear & 25 & 1 & 1/2 & 1.0 & 0.50 & 0.1 \\ 
   18 & nonlinear & linear & 25 & 1 & 1/2 & 0.0 & 0.50 & 0.1 \\ 
   19 & nonlinear & linear & 25 & 1 & 1/2 & 1.0 & 0.50 & 0.1 \\ 
   20 & nonlinear & nonlinear & 25 & 1 & 1/2 & 0.0 & 0.50 & 0.1 \\ 
   21 & nonlinear & nonlinear & 25 & 1 & 1/2 & 1.0 & 0.50 & 0.1 \\  \addlinespace\midrule\addlinespace 
   22 & linear & linear & 25 & 1 & 1/2 & 0.5 & 0.08 & 0.1 \\ 
   23 & linear & linear & 25 & 1 & 1/2 & 0.5 & cov.-dep. & 0.1 \\ 
   24 & nonlinear & linear & 25 & 1 & 1/2 & 0.5 & 0.08 & 0.1 \\ 
   25 & nonlinear & linear & 25 & 1 & 1/2 & 0.5 & cov.-dep. & 0.1 \\ 
   26 & nonlinear & nonlinear & 25 & 1 & 1/2 & 0.5 & 0.08 & 0.1 \\ 
   27 & nonlinear & nonlinear & 25 & 1 & 1/2 & 0.5 & cov.-dep. & 0.1 \\  \addlinespace\midrule\addlinespace 
   28 & linear & linear & 25 & 1 & 1/2 & 0.5 & 0.50 & 0.3 \\ 
   29 & linear & linear & 25 & 1 & 1/2 & 0.5 & 0.50 & 1.0 \\ 
   30 & nonlinear & linear & 25 & 1 & 1/2 & 0.5 & 0.50 & 0.3 \\ 
   31 & nonlinear & linear & 25 & 1 & 1/2 & 0.5 & 0.50 & 1.0 \\ 
   32 & nonlinear & nonlinear & 25 & 1 & 1/2 & 0.5 & 0.50 & 0.3 \\ 
   33 & nonlinear & nonlinear & 25 & 1 & 1/2 & 0.5 & 0.50 & 1.0 \\ 
   \hline
\end{tabular}
\end{table}

The impact of different aspects of the data-generating process was investigated by systematic variation of the parameters across subsets of the configurations:
\begin{itemize}
    \item \textbf{Hazard complexity:} Configurations 1--9 investigate the effect of increasing hazard complexity by varying the functions $f_R$ and $f_\tau$ (linear vs. nonlinear), as well as the dimensions $p_R$ and $p_\tau$.
    \item \textbf{Event type distribution:} Configurations 2, 5, 8, and 10--15 investigate the effect of the event type distribution by varying the parameter $\theta$. 
    \item \textbf{Treatment heterogeneity:} Configurations 2, 5, 8, and 16--21 investigate the effect of treatment heterogeneity by varying the parameter $\gamma_1$. 
    \item \textbf{Treatment assignment:} Configurations 2, 5, 8, and 22--27 investigate the effect of the treatment assignment mechanism by varying the propensity score function $e$.
    \item \textbf{Censoring rate:} Configurations 2, 5, 8, and 28--33 investigate the effect of different censoring rates by varying the parameter $\lambda$. The mean censoring rates that were realized across these settings are displayed in \Cref{tab:cens}. 
\end{itemize}

\begin{table}[ht]
\centering
\caption{Mean censoring rates achieved by varying $\lambda$.}
\label{tab:cens}
\begin{tabular}{c|c|c|c|c}
\toprule
Config. & $f_R$ & $f_\tau$ & $\lambda$ & Censoring Rate \\
\midrule
2 & linear & linear & 0.1 & 0.213 \\
5 & linear & linear & 0.3 & 0.366 \\
8 &linear & linear & 1.0 & 0.511 \\
28 & nonlinear & linear & 0.1 & 0.096 \\
29 & nonlinear & linear & 0.3 & 0.195 \\
30 & nonlinear & linear& 1.0& 0.338 \\
31 &nonlinear & nonlinear & 0.1 & 0.099 \\
32 &nonlinear & nonlinear& 0.3 & 0.204 \\
33 & nonlinear & nonlinear& 1.0 & 0.357 \\
\bottomrule
\end{tabular}
\end{table}

\newpage

\section{Complementary Simulation Results} \label{sec:A.further}
This section provides additional figures and tables complementing the simulation study presented in Section \ref{sec:5}. \Cref{tab:acctau} reports, for each meta-learner and simulation configuration, the percentage of runs that resulted in constant CATE models. Constant models occur most frequently in complex settings with nonlinear $f_R$ and $f_\tau$, as well as under strong treatment imbalance, particularly for R-Learners as well as DR- and M-Learners employing elastic net. 

\begin{figure}
    \centering
    \includegraphics[width=0.9\linewidth]{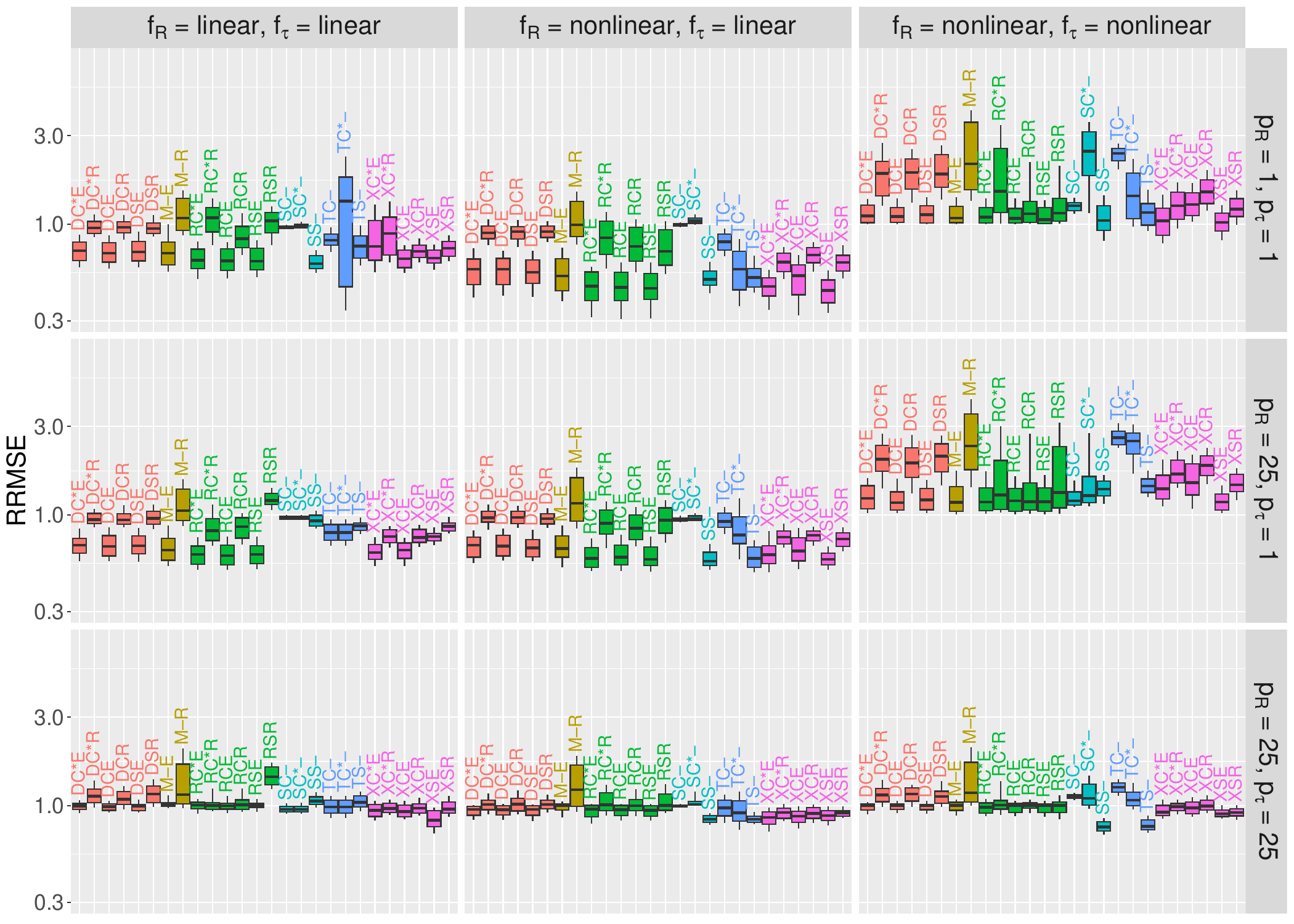}
    \caption{Boxplots showing the RRMSE of all meta-learners for varying complexity of the hazard functions. 
    The parameters $\theta$ and $\gamma_1$ are both consistently set to 0.5 while $e(\boldsymbol{Z}) \equiv 0.5$ and $\lambda=0.1$.}
    \label{fig:51aRRMSE}
\end{figure}

\begin{figure}
    \centering
    \includegraphics[width=0.9\linewidth]{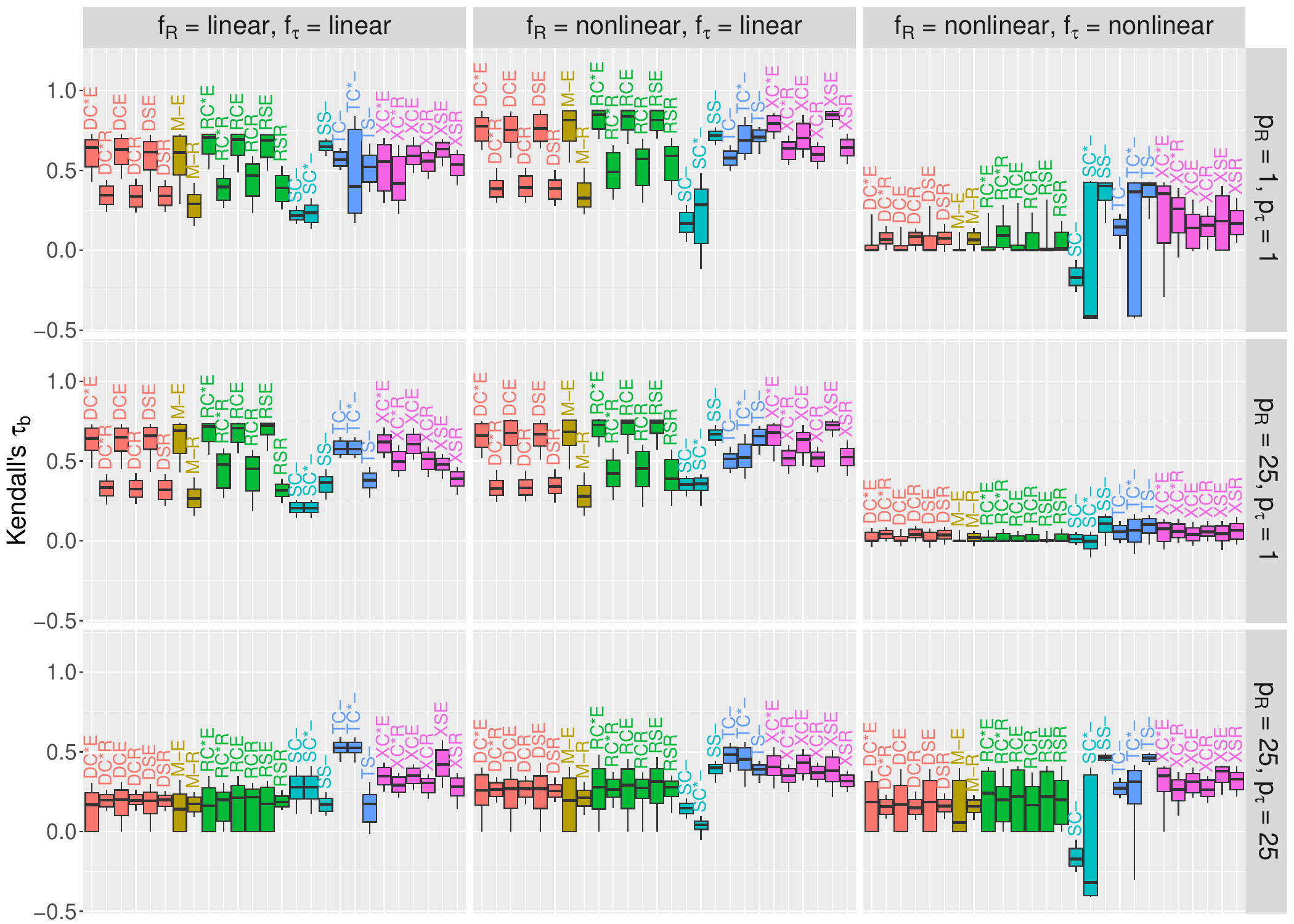}
    \caption{Boxplots showing $\tau_b$ of all meta-learners for varying complexity of the hazard functions. 
    The parameters $\theta$ and $\gamma_1$ are both consistently set to 0.5 while $e(\boldsymbol{Z}) \equiv 0.5$ and $\lambda=0.1$.}
    \label{fig:51aTAU}
\end{figure}

\begin{figure}
    \centering
    \includegraphics[width=0.9\linewidth]{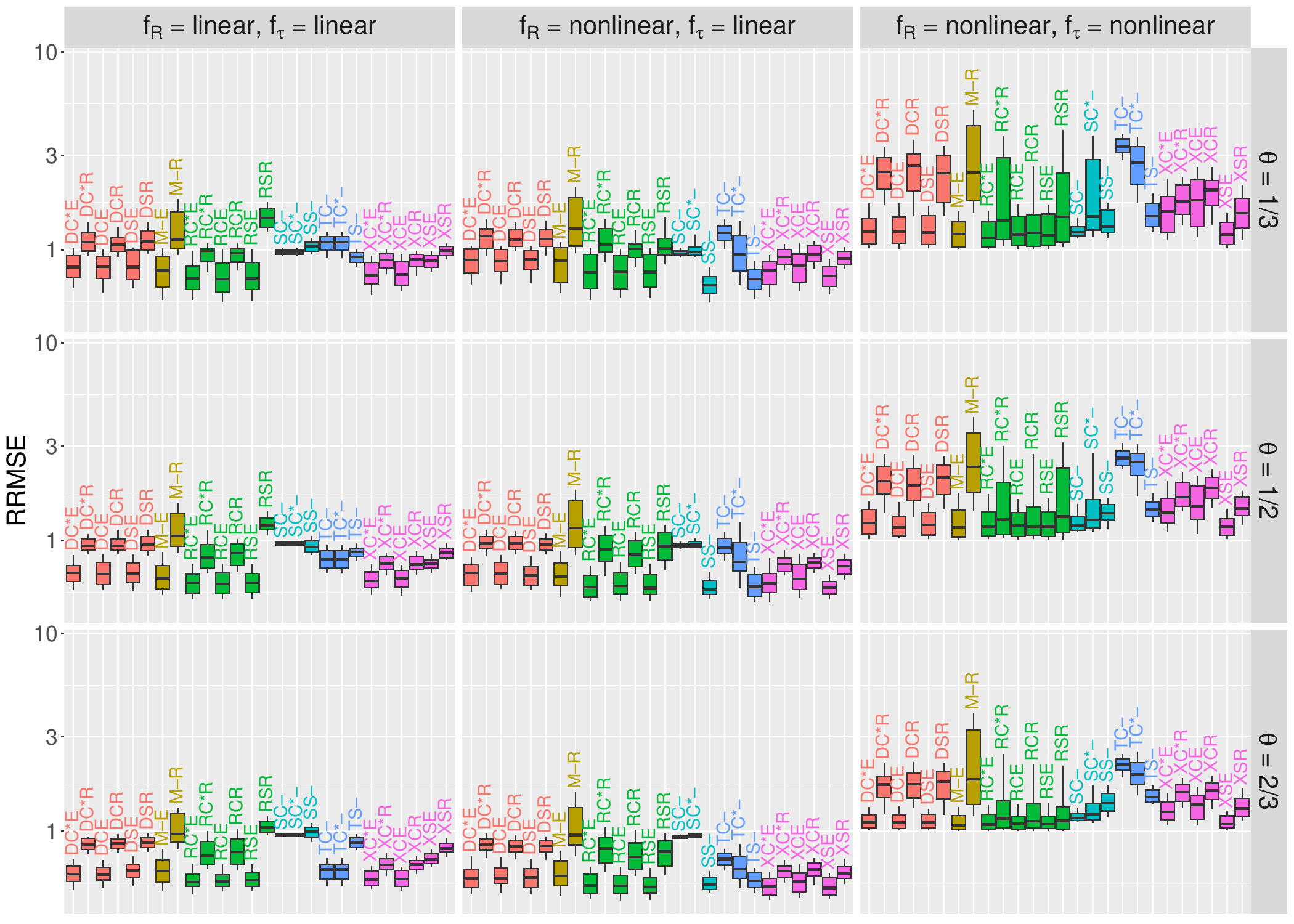}
    \caption{Boxplots showing the RRMSE of all meta-learners for varying event type distribution, where $\theta$ corresponds to the share of expected type 1 events before censoring is applied. 
    The remaining parameters are set according to $p_R=p_\tau=1$, $\gamma_1 = 0.5$, $e(\boldsymbol{Z}) \equiv 0.5$ and $\lambda=0.1$.}
    \label{fig:52aRRMSE}
\end{figure}

\begin{figure}
    \centering
    \includegraphics[width=0.9\linewidth]{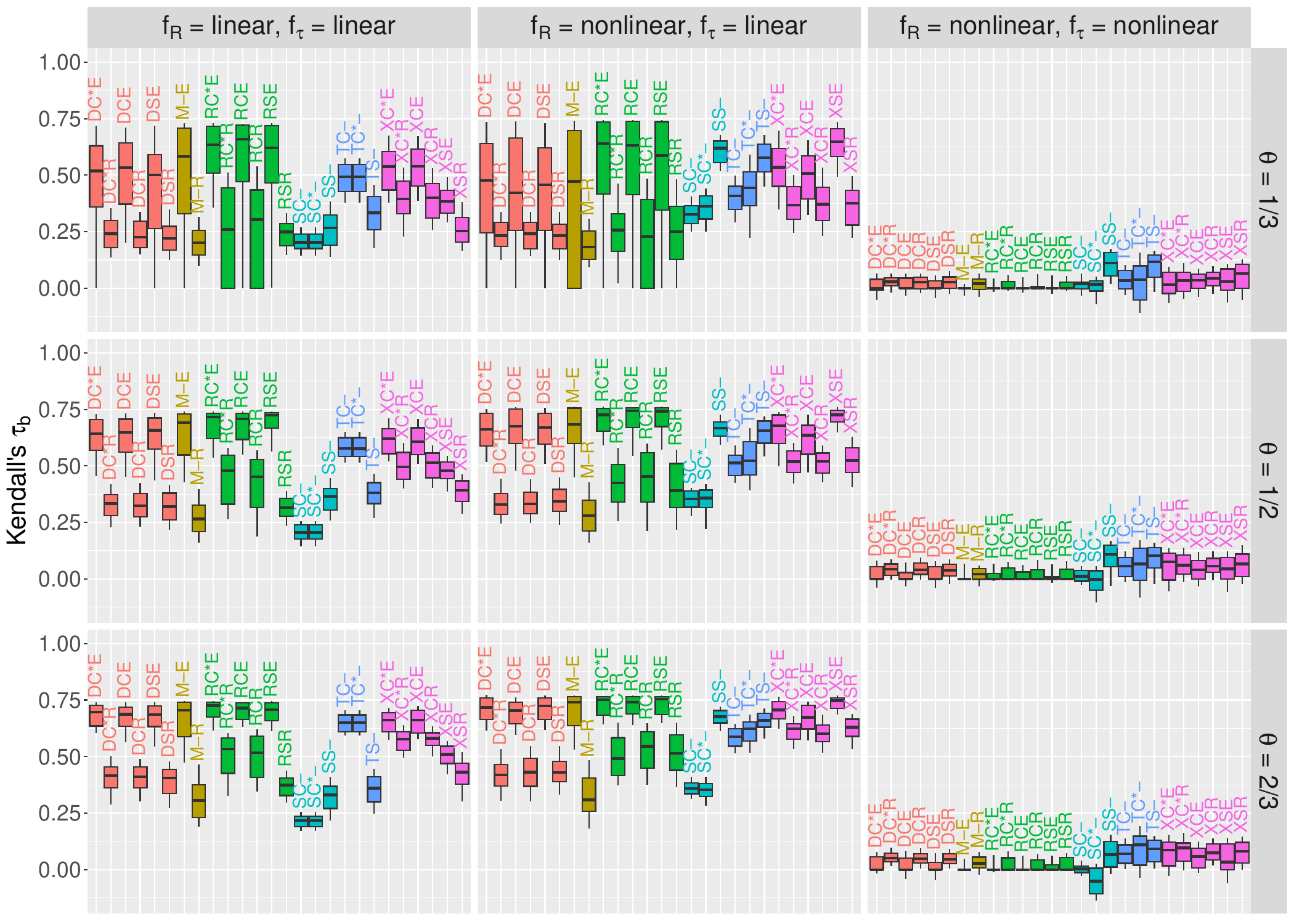}
    \caption{Boxplots showing $\tau_b$ of all meta-learners for varying event type distribution, where $\theta$ corresponds to the share of expected type 1 events before censoring is applied. 
    The remaining parameters are set according to $p_R=p_\tau=1$, $\gamma_1 = 0.5$, $e(\boldsymbol{Z}) \equiv 0.5$ and $\lambda=0.1$.}
    \label{fig:52aTAU}
\end{figure}

\begin{figure}
    \centering
    \includegraphics[width=0.9\linewidth]{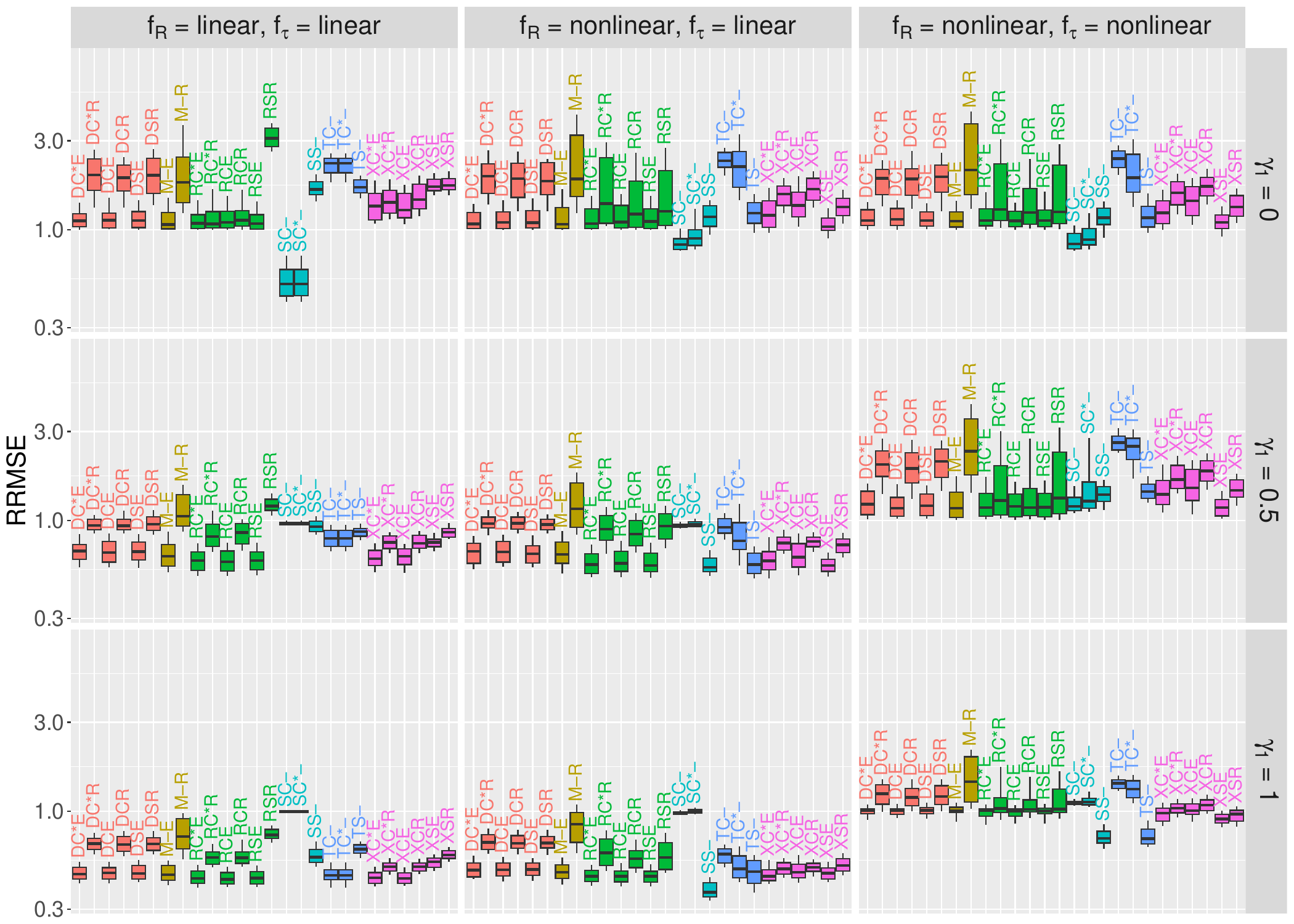}
    \caption{Boxplots showing the RRMSE of all meta-learners for varying strength of treatment heterogeneity. A higher positive value of $\gamma_1$ indicates higher treatment heterogeneity. 
    The remaining parameters are set according to $p_R=p_\tau=1$, $\theta= 0.5$, $e(\boldsymbol{Z}) \equiv 0.5$ and $\lambda=0.1$.}
    \label{fig:53aRRMSE}
\end{figure}

\begin{figure}
    \centering
    \includegraphics[width=0.9\linewidth]{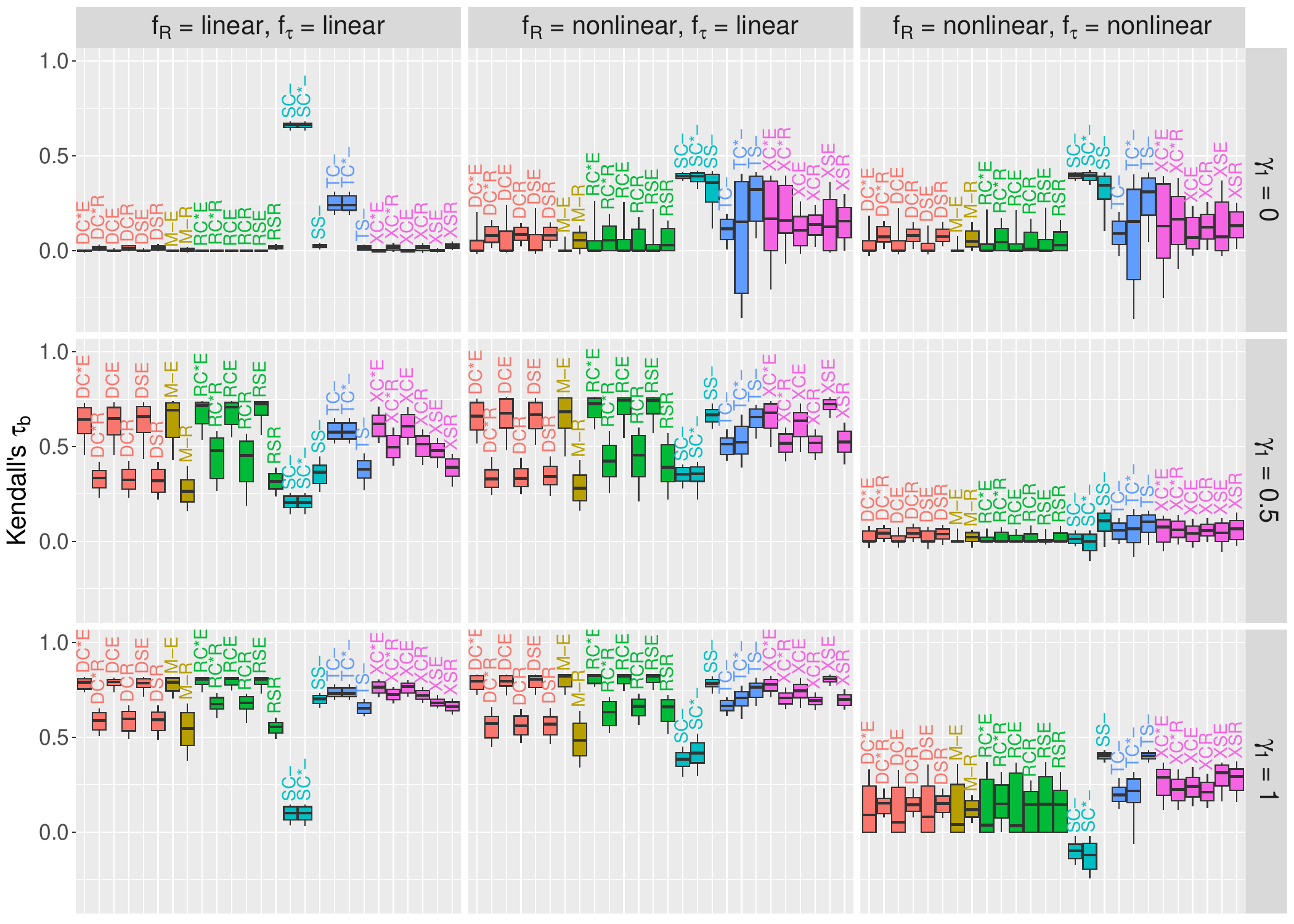}
    \caption{Boxplots showing $\tau_b$ of all meta-learners for varying strength of treatment heterogeneity. A higher positive value of $\gamma_1$ indicates higher treatment heterogeneity. 
    The remaining parameters are set according to $p_R=p_\tau=1$, $\theta= 0.5$, $e(\boldsymbol{Z}) \equiv 0.5$ and $\lambda=0.1$.}
    \label{fig:53aTAU}
\end{figure}

\begin{figure}
    \centering
    \includegraphics[width=0.9\linewidth]{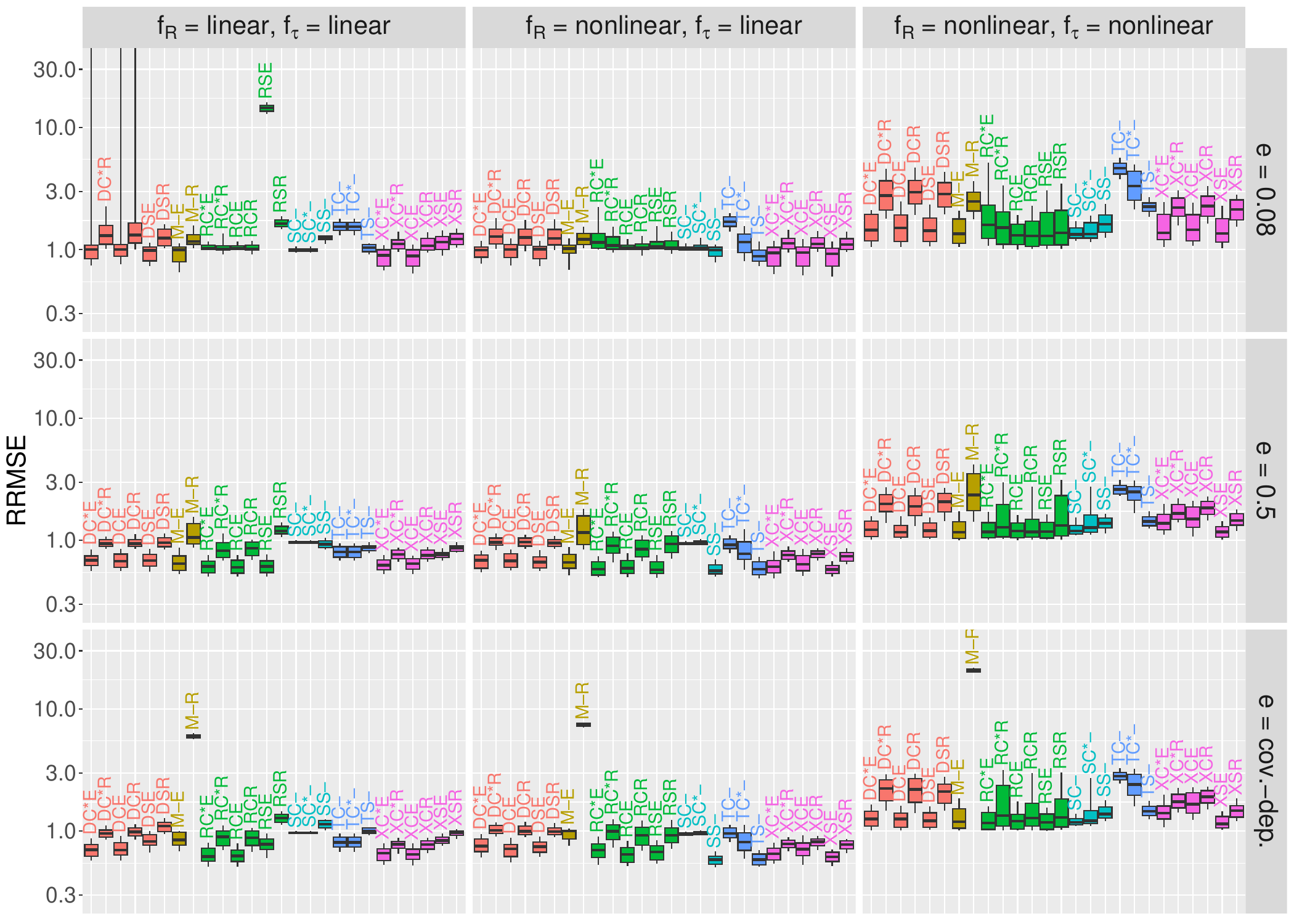}
    \caption{Boxplots showing the RRMSE of all meta-learners for varying treatment assignment, expressed by the propensity score $e$. 
    The remaining parameters are set according to $p_R=p_\tau=1$ and $\theta= \gamma_1=0.5$ and $\lambda=0.1$.}
    \label{fig:54aRRMSE}
\end{figure}

\begin{figure}
    \centering
    \includegraphics[width=0.9\linewidth]{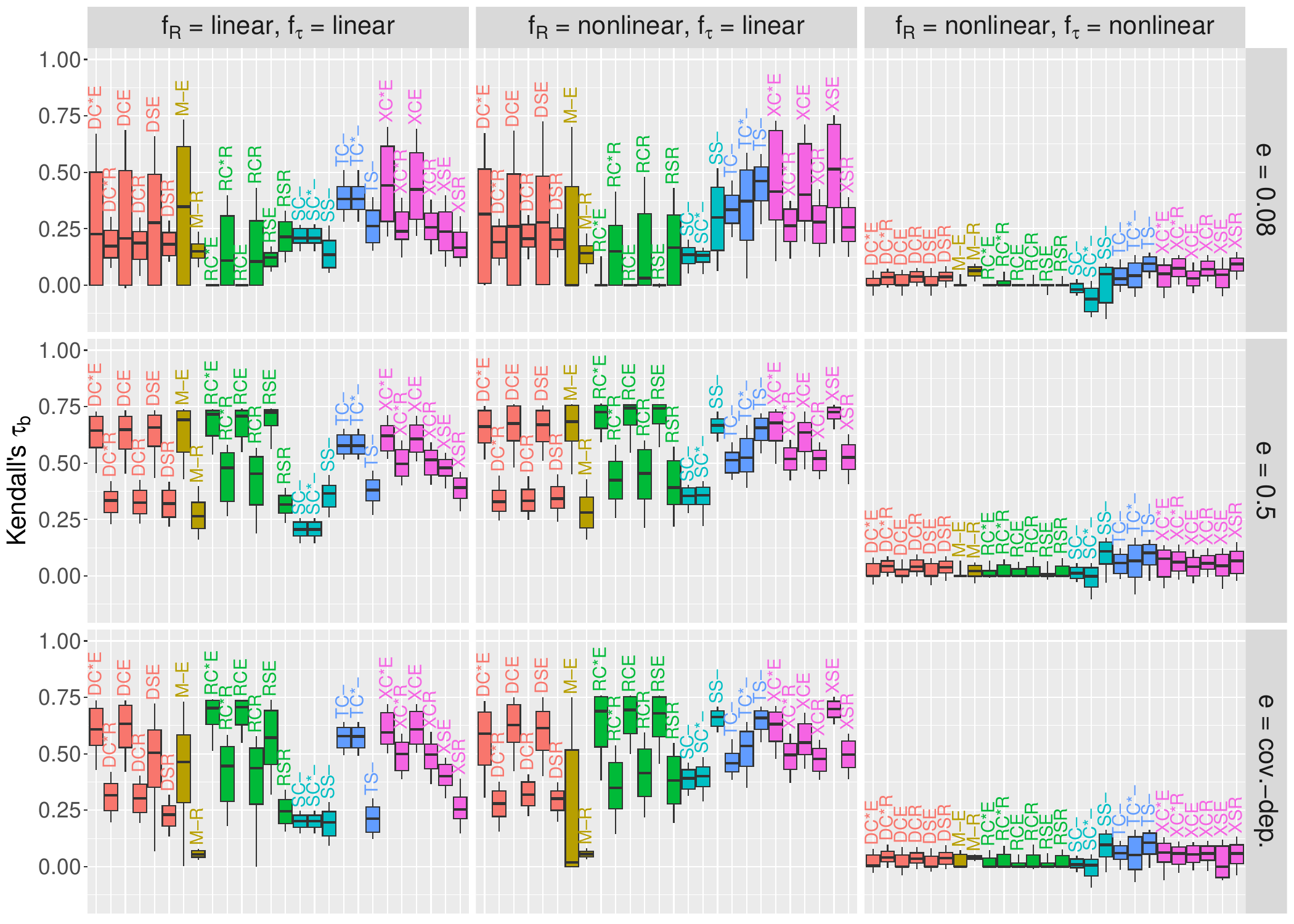}
    \caption{Boxplots showing $\tau_b$ of all meta-learners for varying treatment assignment, expressed by the propensity score $e$. 
    The remaining parameters are set according to $p_R=p_\tau=1$ and $\theta= \gamma_1=0.5$ and $\lambda=0.1$.}
    \label{fig:54aTAU}
\end{figure}

\begin{figure}
    \centering
    \includegraphics[width=0.9\linewidth]{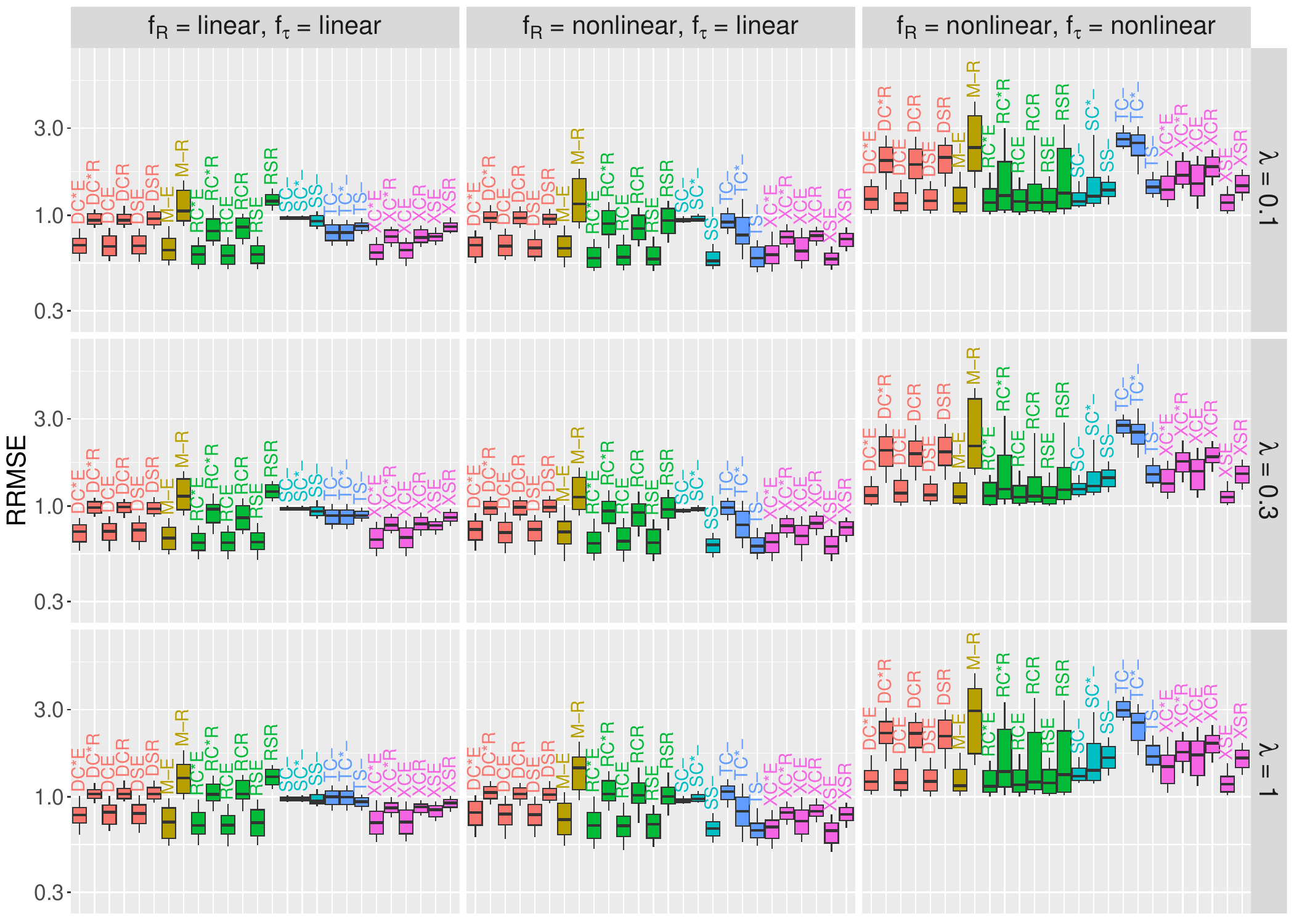}
    \caption{Boxplots showing the RRMSE of all meta-learners for varying censoring distributions through scale parameter $\lambda$. An increase in $\lambda$ corresponds to an increase in the censoring rate. 
    The remaining parameters are set according to $p_R=p_\tau=1$ and $\theta= \gamma_1=0.5$ and $e(\boldsymbol{Z})\equiv 0.5$.}
    \label{fig:55aRRMSE}
\end{figure}

\begin{figure}
    \centering
    \includegraphics[width=0.9\linewidth]{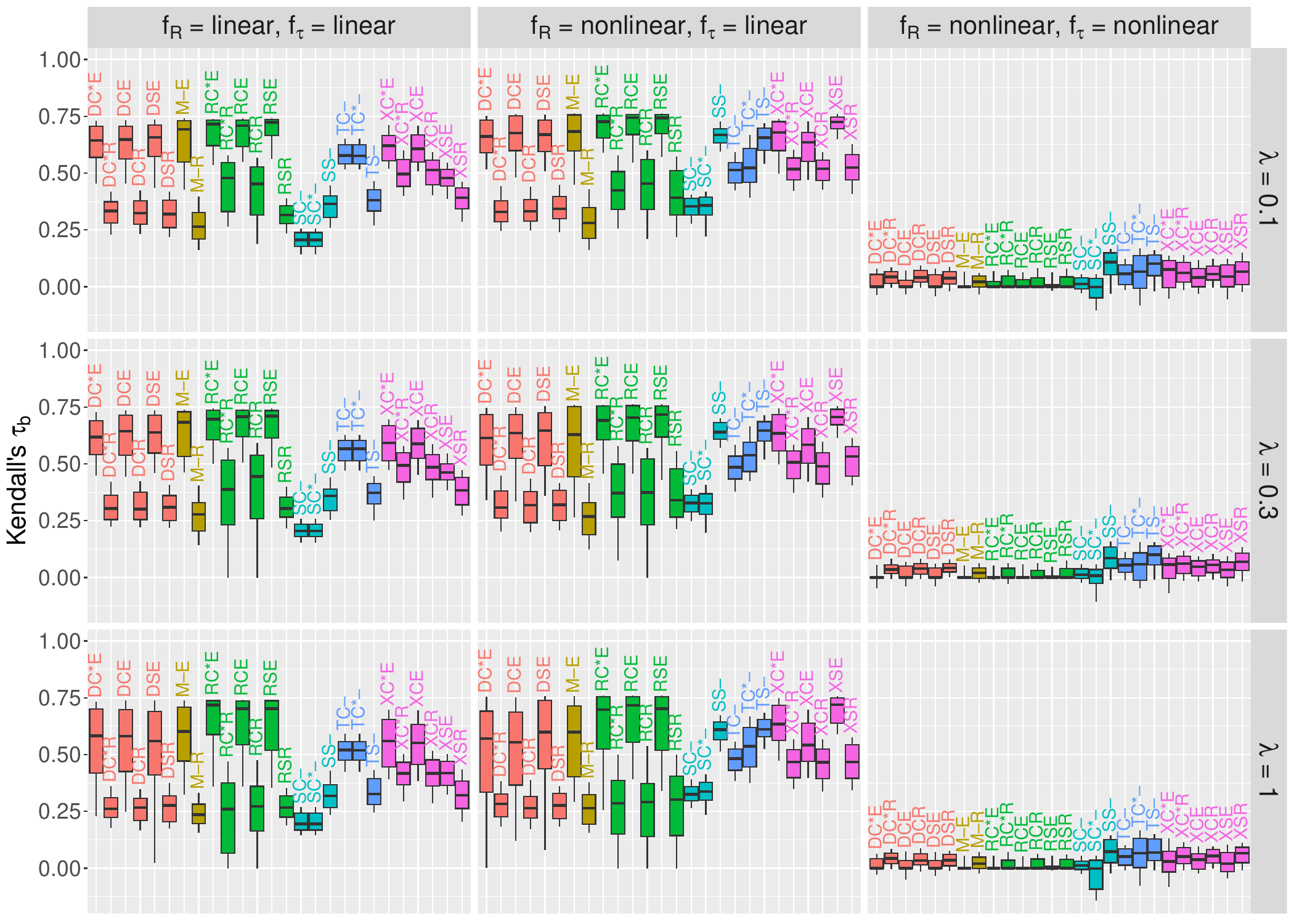}
    \caption{Boxplots showing $\tau_b$ of all meta-learners for varying censoring distributions through scale parameter $\lambda$. An increase in $\lambda$ corresponds to an increase in the censoring rate.  
    The remaining parameters are set according to $p_R=p_\tau=1$ and $\theta= \gamma_1=0.5$ and $e(\boldsymbol{Z})\equiv 0.5$.}
    \label{fig:55aTAU}
\end{figure}

Subsequently, Figures \ref{fig:51aRRMSE}--\ref{fig:55aTAU} show boxplots of the performances of all considered meta-learners in terms of RRMSE and $\tau_b$, including those omitted from the figures in the main text for clarity, across all simulation settings.

\begin{table}[h]
\caption{Amount of simulation runs in $\%$ in which constant CATE models were obtained for each meta-learner and configuration. Learners that did not produce any constant models are excluded.}
\label{tab:acctau}
\resizebox{\textwidth}{!}{
    \begin{tabular}{c|rrrrrrrrrrrrrrrrrrr}
Config. & DC*E & DC*R & DCE & DCR & DSE & DSR & M--E & M--R & RC*E & RC*R & RCE & RCR & RSE & RSR & XC*E & XC*R & XCE & XCR & XSE\\
\midrule
1 & 2 & 1 & 1 & 0 & 2 & 0 & 7 & 1 & 0 & 0 & 0 & 7 & 0 & 1 & 0 & 0 & 0 & 0 & 0\\
2 & 0 & 0 & 0 & 0 & 2 & 0 & 4 & 2 & 0 & 4 & 1 & 8 & 0 & 0 & 0 & 0 & 0 & 0 & 0\\
3 & 23 & 0 & 15 & 0 & 14 & 0 & 39 & 1 & 34 & 23 & 26 & 30 & 31 & 0 & 0 & 0 & 0 & 0 & 0\\
4 & 0 & 0 & 0 & 0 & 0 & 0 & 1 & 1 & 0 & 2 & 0 & 7 & 0 & 4 & 0 & 0 & 0 & 0 & 0\\
5 & 2 & 0 & 0 & 0 & 1 & 0 & 4 & 1 & 0 & 4 & 0 & 9 & 0 & 6 & 0 & 0 & 0 & 0 & 0\\
6 & 15 & 0 & 13 & 0 & 14 & 0 & 36 & 3 & 21 & 7 & 24 & 15 & 18 & 8 & 0 & 0 & 0 & 0 & 1\\
7 & 36 & 0 & 45 & 0 & 30 & 0 & 73 & 1 & 64 & 13 & 58 & 48 & 63 & 41 & 0 & 0 & 5 & 0 & 17\\
8 & 35 & 0 & 43 & 0 & 32 & 0 & 67 & 1 & 58 & 41 & 55 & 57 & 61 & 44 & 2 & 0 & 3 & 0 & 10\\
9 & 26 & 0 & 30 & 0 & 29 & 0 & 43 & 2 & 32 & 23 & 34 & 32 & 38 & 23 & 4 & 0 & 0 & 0 & 3\\
 \addlinespace\midrule\addlinespace 
10 & 12 & 0 & 6 & 0 & 18 & 0 & 19 & 3 & 5 & 33 & 11 & 28 & 10 & 0 & 0 & 0 & 0 & 0 & 0\\
11 & 0 & 0 & 0 & 0 & 0 & 0 & 2 & 2 & 0 & 2 & 0 & 3 & 0 & 0 & 0 & 0 & 0 & 0 & 0\\
12 & 10 & 0 & 16 & 0 & 10 & 0 & 27 & 2 & 14 & 10 & 15 & 31 & 16 & 20 & 0 & 1 & 0 & 0 & 0\\
13 & 0 & 0 & 0 & 0 & 0 & 0 & 5 & 1 & 0 & 0 & 0 & 2 & 0 & 1 & 0 & 0 & 0 & 0 & 0\\
14 & 34 & 0 & 40 & 0 & 41 & 0 & 72 & 1 & 68 & 39 & 66 & 63 & 61 & 45 & 3 & 0 & 2 & 0 & 13\\
15 & 43 & 0 & 33 & 0 & 46 & 0 & 74 & 1 & 67 & 46 & 65 & 58 & 63 & 52 & 4 & 0 & 3 & 0 & 13\\
 \addlinespace\midrule\addlinespace 
16 & 39 & 0 & 46 & 0 & 40 & 0 & 73 & 2 & 66 & 62 & 63 & 56 & 62 & 0 & 1 & 0 & 5 & 1 & 0\\
17 & 0 & 0 & 0 & 0 & 0 & 0 & 0 & 1 & 0 & 0 & 0 & 0 & 0 & 0 & 0 & 0 & 0 & 0 & 0\\
18 & 45 & 0 & 39 & 0 & 34 & 0 & 71 & 3 & 58 & 28 & 53 & 47 & 61 & 42 & 5 & 0 & 1 & 0 & 16\\
19 & 0 & 0 & 0 & 0 & 0 & 0 & 0 & 2 & 0 & 0 & 0 & 0 & 0 & 0 & 0 & 0 & 0 & 0 & 0\\
20 & 32 & 0 & 31 & 0 & 39 & 0 & 73 & 1 & 56 & 33 & 59 & 42 & 55 & 39 & 1 & 0 & 4 & 0 & 11\\
21 & 25 & 0 & 29 & 0 & 26 & 0 & 45 & 0 & 48 & 20 & 47 & 30 & 41 & 26 & 0 & 0 & 0 & 0 & 4\\
 \addlinespace\midrule\addlinespace 
22 & 20 & 0 & 19 & 1 & 15 & 0 & 32 & 2 & 93 & 45 & 92 & 43 & 0 & 0 & 0 & 0 & 0 & 0 & 0\\
23 & 2 & 0 & 3 & 0 & 6 & 0 & 10 & 0 & 1 & 8 & 2 & 15 & 3 & 0 & 0 & 0 & 0 & 0 & 0\\
24 & 20 & 0 & 24 & 0 & 22 & 0 & 48 & 3 & 79 & 33 & 97 & 49 & 74 & 34 & 0 & 0 & 0 & 0 & 0\\
25 & 6 & 0 & 0 & 1 & 2 & 1 & 46 & 0 & 7 & 9 & 0 & 6 & 4 & 9 & 0 & 0 & 0 & 0 & 0\\
26 & 36 & 0 & 43 & 0 & 43 & 0 & 71 & 3 & 83 & 59 & 99 & 80 & 81 & 68 & 1 & 0 & 0 & 0 & 2\\
27 & 30 & 0 & 35 & 0 & 37 & 0 & 56 & 0 & 60 & 38 & 58 & 45 & 64 & 44 & 0 & 0 & 0 & 0 & 0\\
 \addlinespace\midrule\addlinespace 
28 & 1 & 1 & 3 & 0 & 1 & 0 & 4 & 3 & 0 & 18 & 0 & 17 & 0 & 0 & 0 & 0 & 0 & 0 & 0\\
29 & 4 & 0 & 5 & 0 & 5 & 0 & 4 & 0 & 5 & 24 & 3 & 19 & 5 & 0 & 0 & 0 & 0 & 0 & 0\\
30 & 1 & 0 & 1 & 0 & 3 & 0 & 8 & 1 & 1 & 10 & 1 & 17 & 3 & 5 & 0 & 0 & 0 & 0 & 0\\
31 & 8 & 0 & 8 & 0 & 7 & 0 & 6 & 2 & 8 & 21 & 5 & 22 & 6 & 20 & 0 & 0 & 0 & 0 & 0\\
32 & 54 & 0 & 37 & 0 & 40 & 0 & 65 & 1 & 65 & 43 & 65 & 52 & 63 & 43 & 2 & 0 & 2 & 0 & 17\\
33 & 37 & 0 & 39 & 0 & 40 & 0 & 68 & 3 & 71 & 43 & 63 & 48 & 61 & 46 & 5 & 0 & 2 & 0 & 15\\
\end{tabular}
}
\end{table}

\newpage

\end{document}